\documentclass[11pt,a4paper]{article}
\pdfoutput=1
\usepackage{jheppub}
\usepackage{subfigure}

\notoc  

\title{Higgs Production via Weak Boson Fusion in the Standard Model and the MSSM}

\author[a,1]{Terrance Figy,%
\note{Former address of the authors, where much
of this work was carried out: 
IPPP, Durham University, Durham DH1~3LE, UK.
}}
\author[b]{Sophy Palmer,}
\author[c]{Georg Weiglein}

\affiliation[a]{CERN, \\ 
CH--1211 Geneva 23, Switzerland}
\affiliation[b]{IThP, KIT, Universit\"at Karlsruhe, \\
76128 Karlsruhe, Germany}
\affiliation[c]{DESY, \\
D--22603 Hamburg, Germany}

\emailAdd{Terrance.Maynard.Figy@cern.ch}
\emailAdd{Sophy.Palmer@particle.uni-karlsruhe.de}
\emailAdd{Georg.Weiglein@desy.de}

\abstract{Weak boson fusion is expected to be an important Higgs production
channel at the LHC. Complete one-loop results for weak boson fusion in the
Standard Model have been obtained by calculating the full virtual electroweak
corrections and photon radiation and implementing these results into the public
Monte Carlo program \tt VBFNLO \rm (which includes the NLO QCD corrections).
Furthermore the dominant supersymmetric one-loop corrections to neutral Higgs
production, in the general case where the MSSM includes complex phases, have
been calculated. These results have been combined with all one-loop corrections
of Standard Model type and with the propagator-type corrections from the Higgs
sector of the MSSM up to the two-loop level. Within the Standard Model the
electroweak corrections are found to be as important as the QCD corrections
after the application of appropriate cuts. The corrections yield a shift in the
cross section of order 5\% for a Higgs of mass 100--200 GeV, confirming the
result obtained previously in the literature. For the production of a light
Higgs boson in the MSSM the Standard Model result is recovered in the decoupling
limit, while the loop contributions from superpartners to the production of
neutral MSSM Higgs bosons can give rise to corrections in excess of 10\% away
from the decoupling region.}


\arxivnumber{1012.4789}

\begin{document}
\input paperdef

\maketitle
\flushbottom


\section{Introduction}
\label{sec:intro}

Weak boson fusion (WBF) is an important Higgs production channel at the
LHC~\cite{09010512,cms,0306109} and at a future Linear Collider~\cite{ilc,clic}.
If the Higgs mechanism is responsible for generating the masses of the weak
gauge bosons $Z$ and $W^{\pm}$, one would expect that at least one Higgs boson
should have a significant coupling to the weak bosons (unless the coupling to
gauge bosons is shared among a large number of Higgs bosons, see
e.g.~\cite{Gunion}) and should therefore be produced in weak boson fusion. 
Besides its r\^{o}le as a discovery channel, it has also been shown that weak
boson fusion production can provide important information on the couplings and
$\cp$-properties of the detected state \cite{0105325,Duehrssi,0406323}. A
precise theoretical prediction of this channel is mandatory in this context.

QCD corrections to weak boson fusion Higgs production at the LHC turned out to
be moderate, at the level of 5\% in the Standard Model (SM), and are
theoretically well under control
\cite{9705337,9206246,0306109,0403297,0403194,08013355,07093513,08014231,
10034451}. Additionally, uncertainties from parton distribution functions (PDFs)
to this channel are quite small \cite{0306109} in the phase space region
relevant for the LHC. In view of the expected accuracies at the LHC
\cite{09010512,cms, Duehrssi} electroweak loop corrections may also be
non-negligible. Besides the relevance of electroweak loop corrections for
reducing the theoretical uncertainties of this channel, they are also of
interest because of the potential effects of new physics entering via
virtual contributions of additional particles in the loops. In the Minimal
Supersymmetric Standard Model (MSSM), the most thoroughly studied extension of
the SM, it has been shown for the case of Higgs production in weak boson fusion
at a future linear collider that supersymmetric (SUSY) loop contributions can
have a sizable impact on the production cross section \cite{0211204}. In
particular, the SUSY loop effects can significantly modify the decoupling
behaviour of the $VVH$ vertex, where $V = Z, W^\pm$, and $H$ is the heavy
$\cp$-even Higgs boson of the MSSM \cite{0211204} (see also
\citere{0204280,0211204} for leading SUSY loop corrections to the production of
the light $\cp$-even Higgs boson in WBF at the Linear Collider; the complete
one-loop contributions to the corresponding process in the SM have been obtained
in \citeres{0211261,0301189,0302198}).

Electroweak loop corrections to Higgs production in WBF at the LHC have recently
received considerable interest. In \citeres{07070381,07104749,HAWK} the full
one-loop electroweak and QCD loop corrections to the total cross section and
differential distributions have been evaluated in the SM. The pure SUSY loop
corrections to this process, without the SM part, have been obtained in
\cite{Michael,10042169}, and the SUSY-QCD corrections have also been
investigated~\cite{9912476}. In the present paper we calculate the complete
electroweak one-loop corrections to Higgs production in WBF at the LHC in the
SM. We furthermore calculate corrections to the production of neutral Higgs
bosons in the MSSM, combining the full one-loop SM-type contributions with the
dominant SUSY one-loop corrections involving the scalar superpartners of the SM
fermions and with the propagator-type corrections up to the two-loop level to
the mass and wavefunction normalisation of the outgoing Higgs boson. For
comparison with the dominant SUSY contributions from sfermions we have also
calculated the full SUSY corrections to the $VVh$ vertex, the weak boson self
energies and the $qqV$ vertices.  We have implemented our results into the
public Monte Carlo program {\tt VBFNLO}~\cite{VBFNLO,VBFNLO2} so that they can be used
in experimental studies. 

Our results go beyond the existing results in the literature in various ways. In
particular, they incorporate loop effects from both SM and SUSY particles.
Additionally, our SUSY loop corrections have been obtained for the general
case of non-vanishing complex phases, which enables an analysis of the possible
impact of $\cp$-violating effects. For comparison, we have furthermore evaluated
the fermion and sfermion loop corrections to the production of the $Z$ boson in
WBF, which is of interest as a potential reference process to which WBF Higgs
production could be calibrated. Where possible, we compare our results with
those available in the literature.

\section{Details of the calculation}
\label{sec:calc}

\subsection{Notations and conventions}

The Higgs sector of the MSSM comprises two scalar doublets, resulting in five
physical Higgs bosons. At lowest order the Higgs sector is $\cp$-conserving,
giving rise to two $\cp$-even states $h$ and $H$, a $\cp$-odd state $A$, and the
charged Higgs bosons $H^\pm$. Besides the gauge couplings, the Higgs sector is
characterised by two independent input parameters, conventionally chosen as
$\MA$ and $\tb$ (in the case of $\cp$-violation one usually chooses $\MHpm$
instead of $\MA$ as the input parameter). Here $\tb$ is the ratio of vacuum
expectation values of the two Higgs doublets. The other Higgs boson masses and
the mixing angle $\alpha$ between the two neutral $\cp$-even states can be
predicted in terms of the input parameters. Higher-order contributions yield
large corrections to the masses and couplings, and can also induce
$\cp$-violation leading to mixing between $h,H$ and $A$ in the case of general
complex SUSY-breaking parameters. The corresponding mass eigenstates are denoted
as $h_1$, $h_2$, $h_3$. 

The superpartners to the left- and right-handed fermions mix, yielding the mass
eigenstates $\tilde f_1,\tilde f_2$. In the off-diagonal entries of the mass
matrix the trilinear couplings $A_f$ and the Higgsino mass parameter $\mu$
enter, which can be complex. Similarly, the mass eigenstates of neutralinos and
charginos need to be determined from matrix diagonalisation, where the
parameters $M_1$ and $M_2$ can be complex. The gluino mass and its phase enter
our results only via the two-loop contributions to the Higgs propagators.

In this work we do not consider corrections due to quark mixing, as these are
expected to be small.

\subsection{Types of corrections}

\begin{figure}[htb!]
     \subfigure[Corrections to the $VVH$ vertex]{
         \label{fig:VVH}
         \resizebox{0.45\hsize}{!}{\includegraphics*{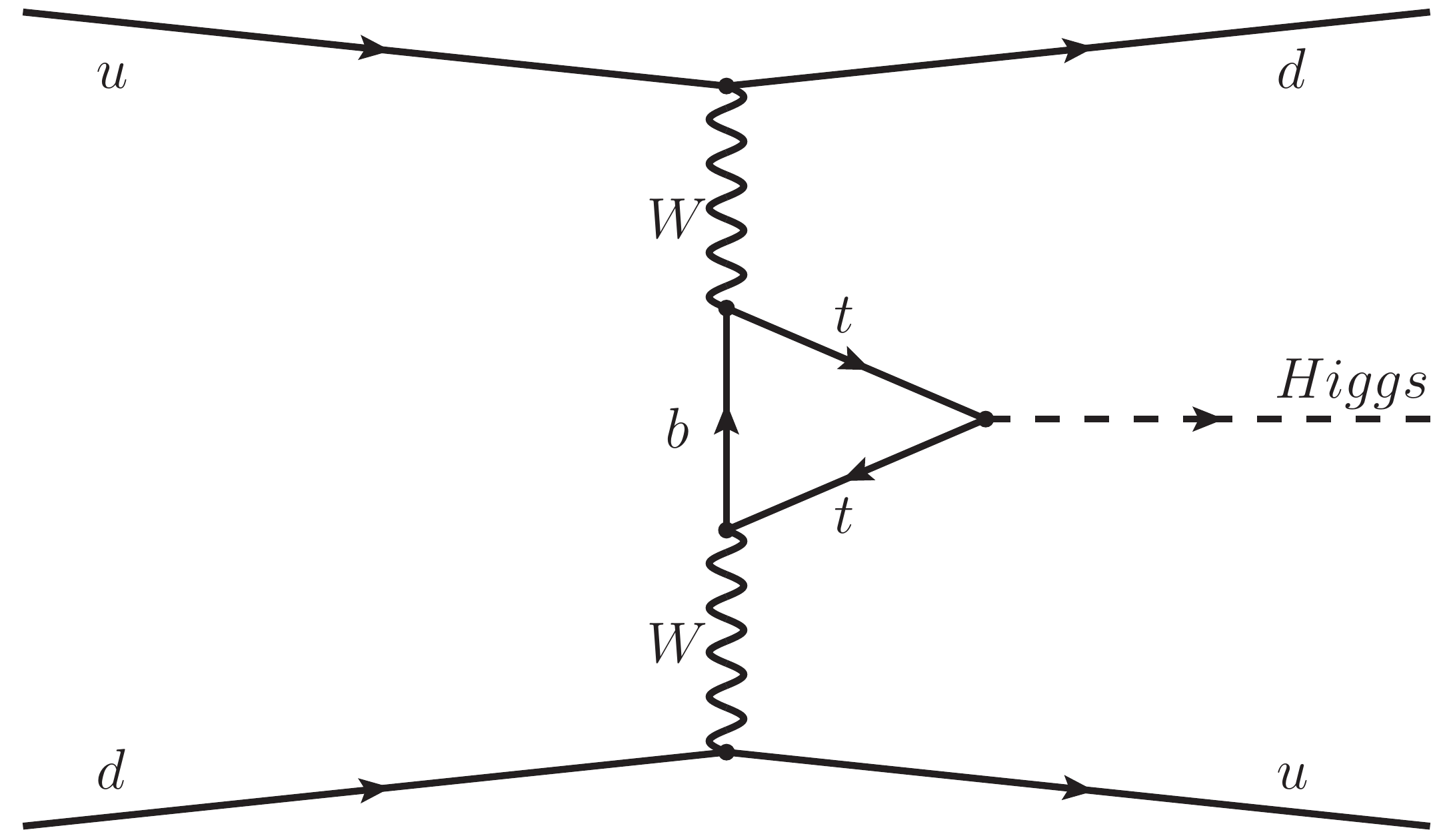}}
         }
     \hspace{.3cm}
     \subfigure[Corrections to the $VV$ self energy]{
         \label{fig:VV}
         \resizebox{0.45\hsize}{!}{\includegraphics*{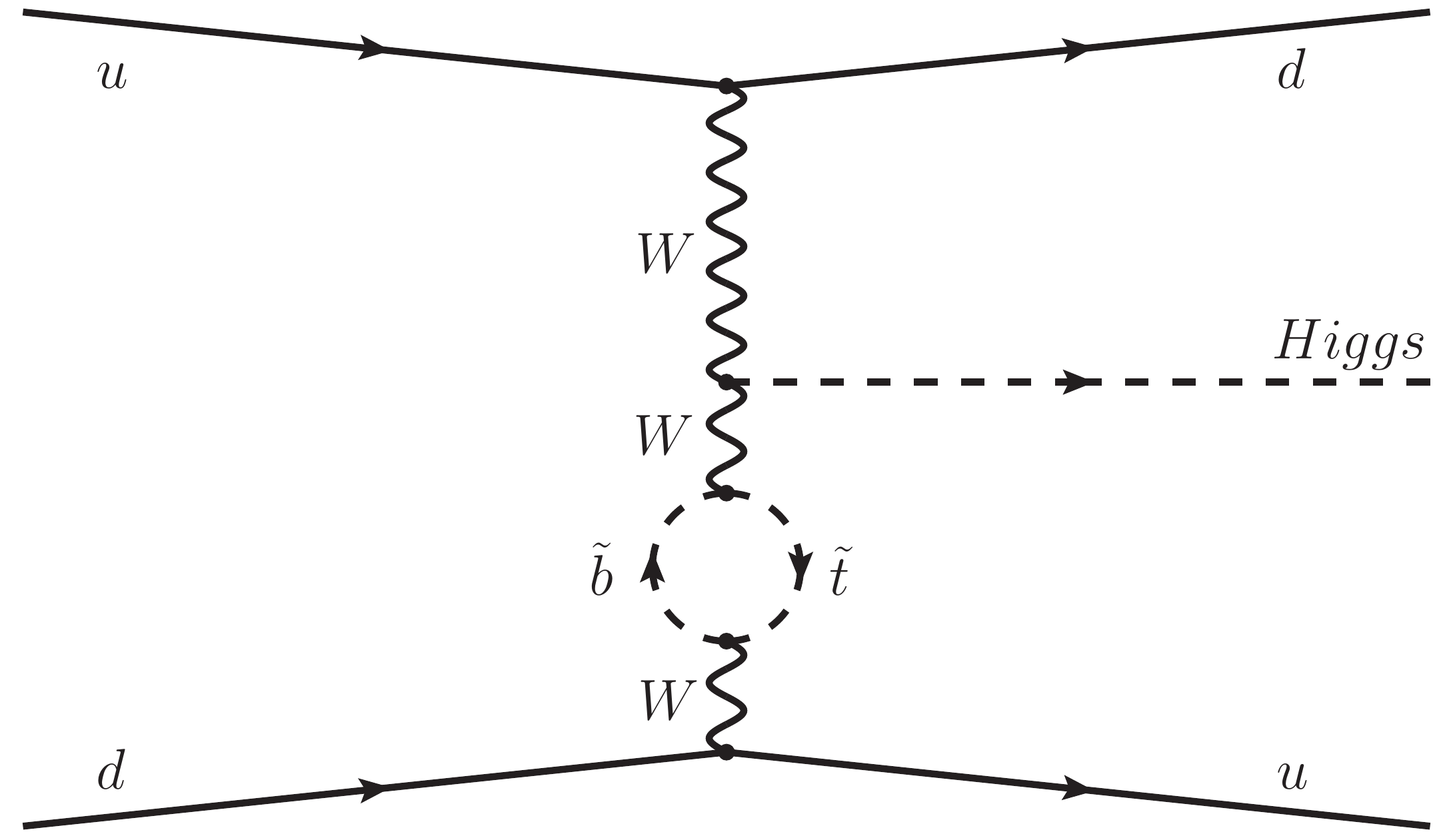}}
         } \\
     \subfigure[Corrections to the $qqV$ vertex]{
         \label{fig:qqV}
         \resizebox{0.45\hsize}{!}{\includegraphics*{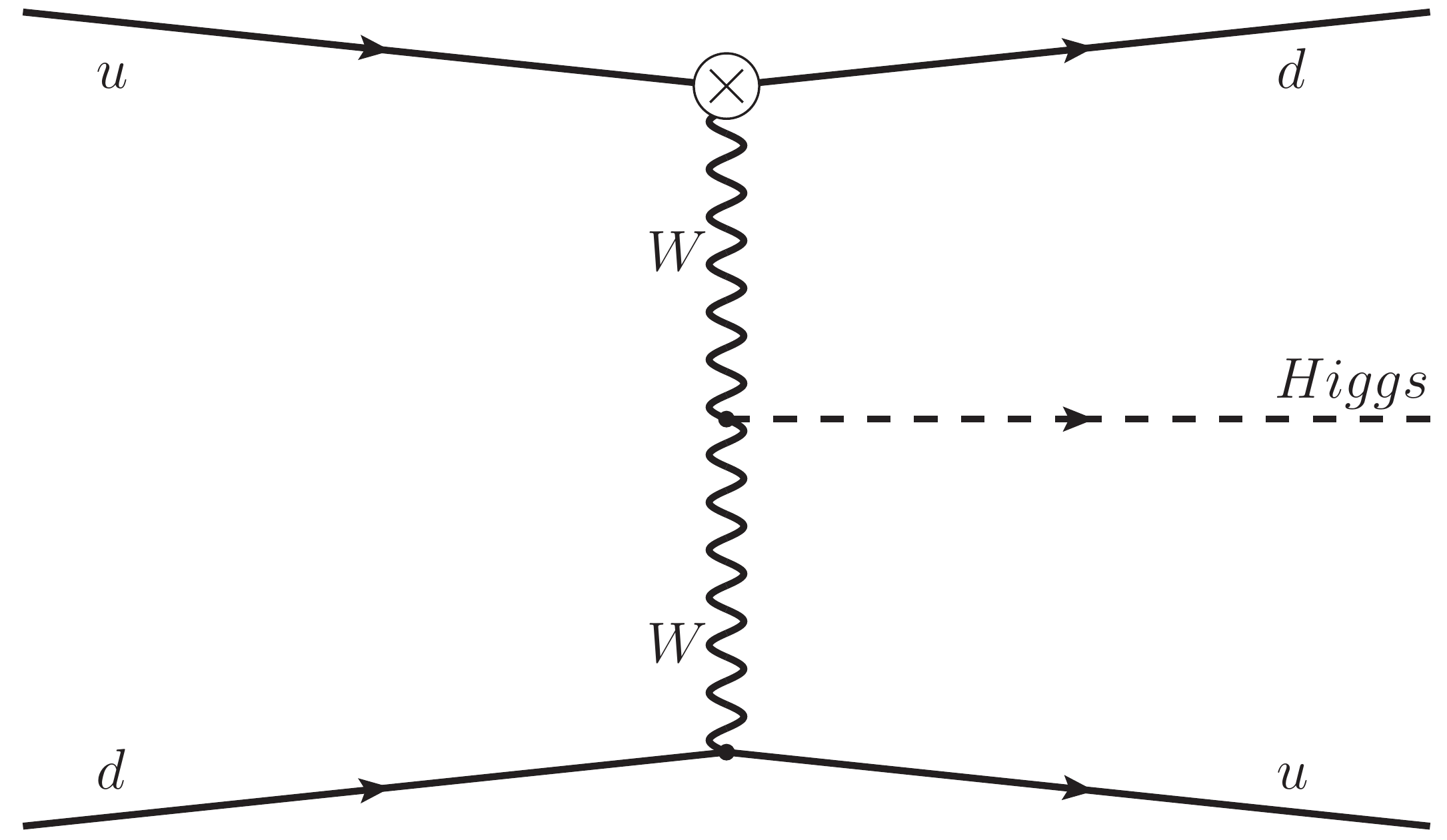}}
         }
     \hspace{.3cm}
     \subfigure[Box diagrams]{
         \label{fig:box}
         \resizebox{0.45\hsize}{!}{\includegraphics*{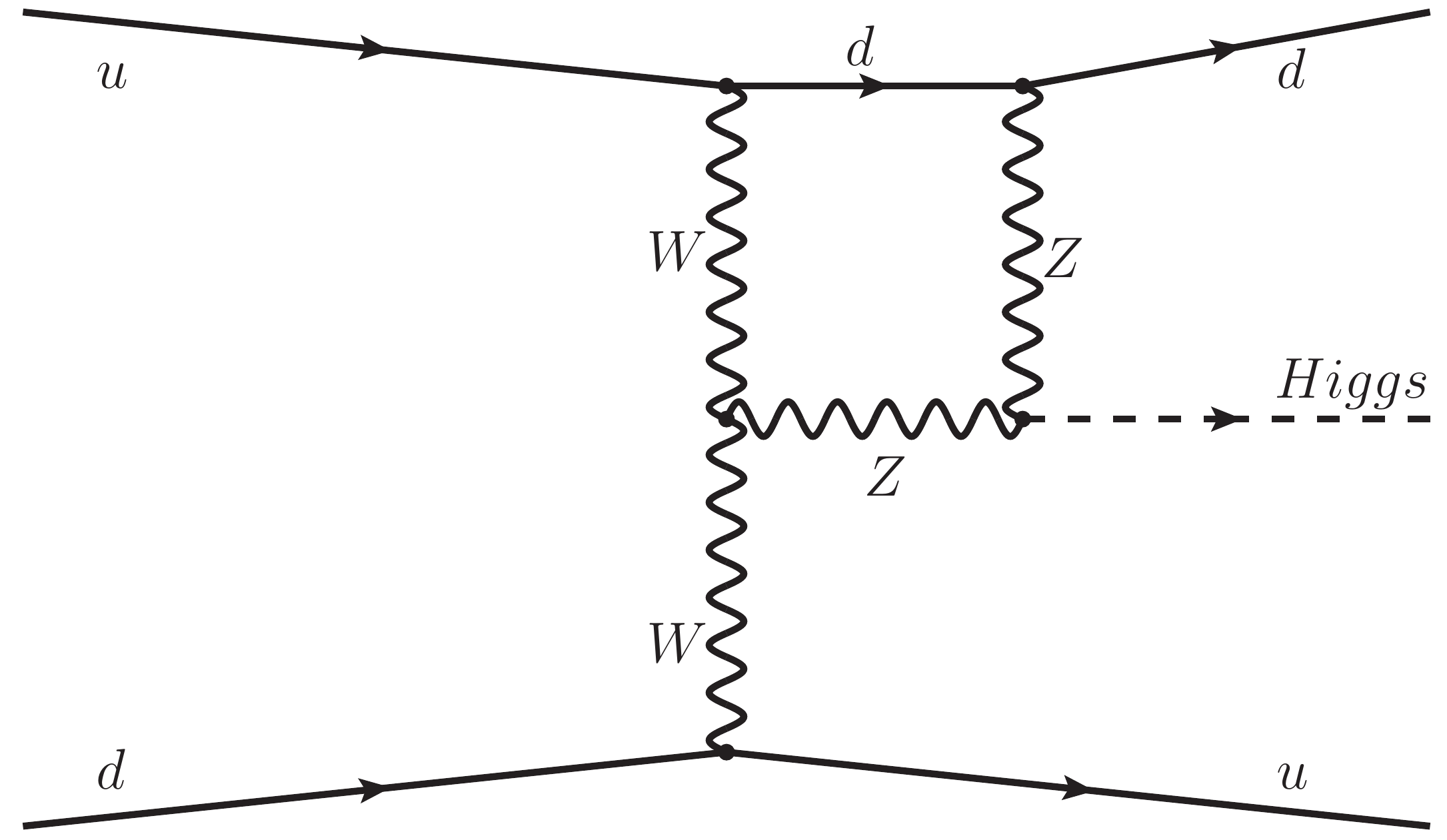}}
         } \\
       \subfigure[Pentagon diagrams]{
         \label{fig:pentagon}
         \resizebox{0.45\hsize}{!}{\includegraphics*{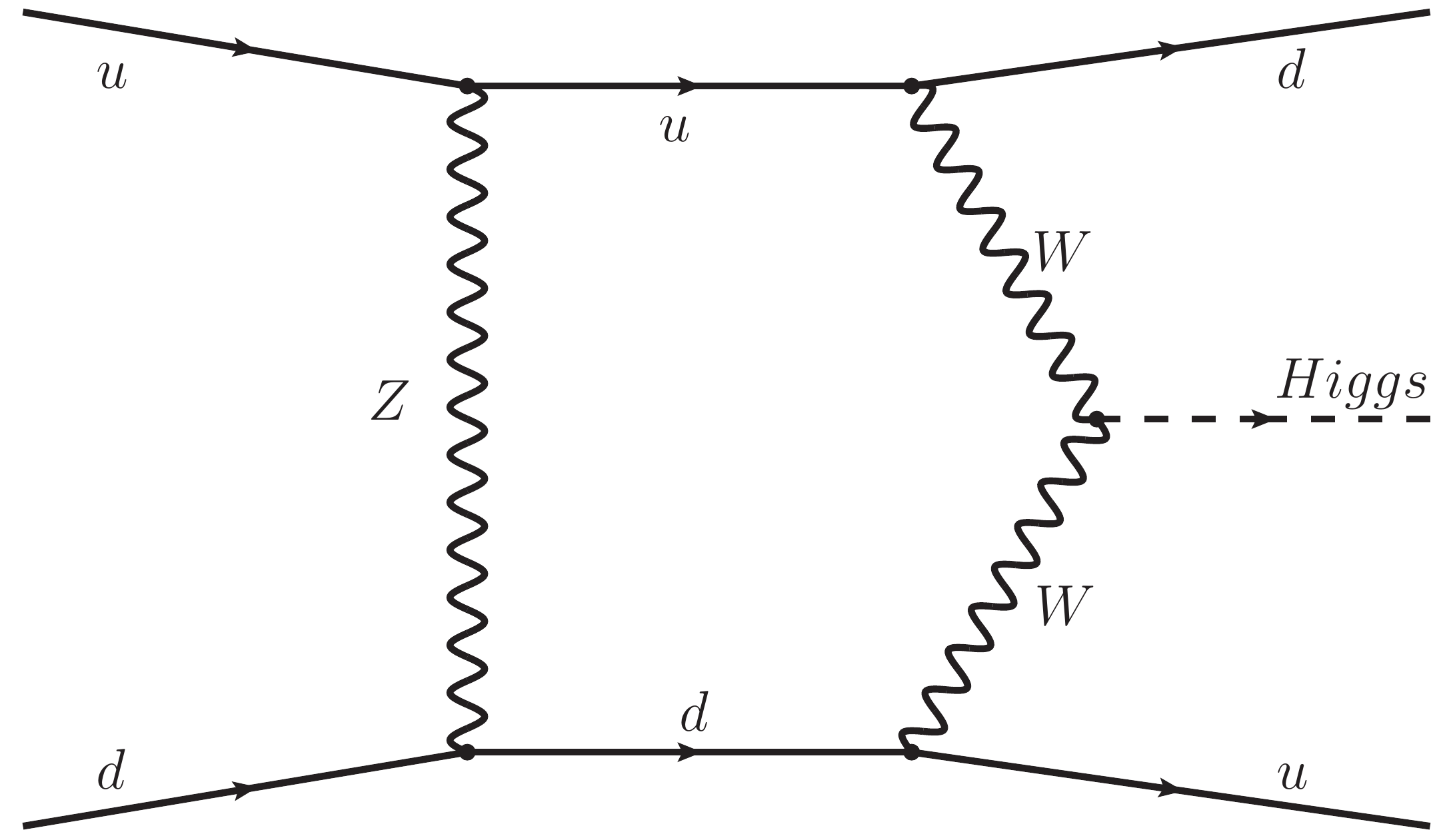}}
         } 
     \hspace{.3cm}
     \subfigure[Real photon emission]{
         \label{fig:photon}
         \resizebox{0.45\hsize}{!}{\includegraphics*{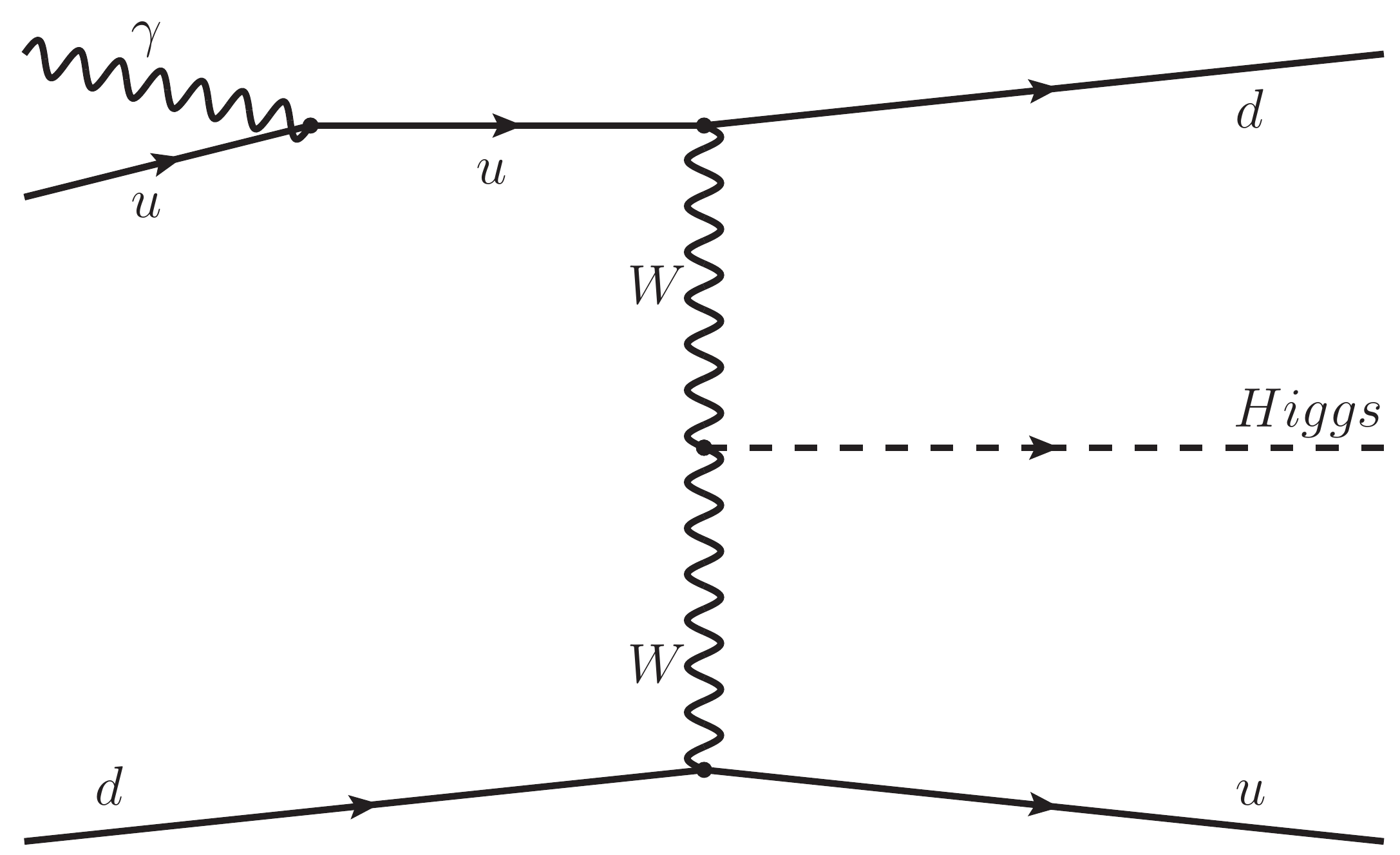}}
         } 
  \vspace{-0.5cm}
     \caption{Types of electroweak corrections to the weak boson fusion
process.} 
     \label{fig:diag_multi}
\end{figure}

In the following we describe details of the calculation of the one-loop
electroweak corrections to WBF production of the SM Higgs boson $H^{\rm SM}$ and
the MSSM Higgs bosons $h$, $H$, $A$ ($h_1$, $h_2$, $h_3$) in the
$\cp$-conserving ($\cp$-violating) case.\footnote{Obviously, at leading order
the $\cp$-odd Higgs $A$ is not produced.} In the SM we take into account the
complete one-loop electroweak corrections to the partonic $2 \to 3$ process,
involving diagrams of pentagon, box, vertex and self-energy type. Generic types
of virtual electroweak one-loop corrections, counterterm contributions and real
photon emission are depicted in \reffi{fig:diag_multi}.  These contributions
have been implemented into the public Monte Carlo program {\tt VBFNLO}
\cite{VBFNLO}, which contains the leading-order result supplemented by the
one-loop QCD corrections in the SM. We therefore obtain results that contain the
full one-loop QCD and electroweak corrections in the SM. In the MSSM we combine
the SM-type contributions (i.e.\ from fermions, gauge bosons and the full MSSM
Higgs sector) with the dominant loop corrections involving the scalar
superpartners of the SM fermions.  We evaluate these contributions for arbitrary
complex phases. Going beyond the sfermion loop corrections, for the $VVh_{i}$
vertex, the $VV$ self energy and the $qqV$ vertices, where $V = W^\pm, Z$ and $i
= 1, 2, 3$, we have obtained the full one-loop contributions from all SUSY
particles. For the propagator corrections to the mass and wavefunction
normalisation of the outgoing Higgs boson, which are known to be sizable in the
MSSM Higgs sector, we incorporate corrections up to the two-loop level as
implemented in the program {\tt
FeynHiggs}~\cite{9812320,9812472,0212020,0611326,07050746,CPC180}. The remaining
SUSY loop corrections (the SUSY-QCD corrections~\cite{9912476}
and the SUSY pentagon and box
diagrams involving the superpartners of the gauge and Higgs bosons) will be
presented in a forthcoming publication.  
The SUSY-QCD corrections have been seen to be small (see the discussion
in~\cite{Michael}) in the SPS scenarios, and are thus expected to be of
sub-leading numerical importance in the scenarios investigated in this work.  We
have calculated the remaining SUSY pentagons and boxes.  In the \Mhmax\
scenario, we explicitly checked that those contributions are numerically
small%
\footnote{In other scenarios -- for instance those with a rather light 
gluino -- these corrections may of course be somwhat larger.}
-- due to this, and the lengthy CPU time needed to evaluate them, these
corrections are not included in this paper or in the public code, but will be
discussed in more detail elsewhere.  On the other hand, the full SM-type 
corrections, which contain for instance possible effects of Sudakov
enhancements at high energies, are included in the results
presented here.

The virtual corrections are supplemented by the diagrams with real photon
emission, see \reffi{fig:photon}. The diagrams with real gluon emission (and the
corresponding virtual corrections) are already included in {\tt VBFNLO}
\cite{VBFNLO}.

In the SM, for the perturbative evaluation of the cross section up to the
one-loop level, it is sufficient to take into account only the contributions of
the squared Born-level matrix element (as well as the one with real photon
emission) and the product of the Born-level and the one-loop amplitude.  In the
MSSM, on the other hand, this is not necessarily the case. Since in
particular for the production of the heavy Higgs bosons the Born-level matrix
element can be very small (or even strictly zero in the case of the $\cp$-odd
Higgs boson), the squared contribution of the fermion/sfermion part of the loop amplitude, $\vert
\mathcal{M}^{\rm loop} \vert ^{2}$, can be numerically relevant and needs to be
incorporated\footnote{Only the squared part of the s/fermion contribution is
considered in order to avoid problems with IR-finiteness.}. The implementation
of our results into \tt VBFNLO \rm is such that, whenever the loop corrections
are a sizeable fraction (greater than 15\%) of the leading order contributions a
warning is output, indicating to the user that the loop squared contributions
will have a significant effect and should be included.
Besides the contributions to Higgs production in WBF we also consider
higher-order corrections to the production of a $Z$ boson in WBF at the partonic
level. This process could in principle be of interest as a reference process
(with a similar signature and, for a light Higgs, similar kinematics) to which
Higgs production could be calibrated (see e.g.\ \cite{ex0502009} for a
discussion), although its experimental feasibility remains questionable at
present~\cite{10014357}. We have calculated the fermion and sfermion loop
corrections to this process (which is also incorporated, including one-loop QCD
corrections, in {\tt VBFNLO} \cite{VBFNLO}), which allows us to compare the
pattern of the radiative corrections for the two processes.

The calculations of the Feynman diagrams were performed using the programs {\tt
FeynArts} \cite{Kublbeck:1990xc,Denner:1992vza,FAorig,FeynArts1,FeynArts2} and {\tt FormCalc}
\cite{FormCalc1,FormCalc2,FormCalc3}, and the loop integrals were evaluated
using {\tt LoopTools} \cite{LoopTools}.  Throughout, we use Dimensional
Reduction (DRED \cite{DRED,DRED2,0503129}).

\subsubsection {Virtual corrections}
\label{sec:virt}

The virtual corrections can be grouped into five different categories:
corrections to the $VVH$ vertex, corrections to the $qqV$ vertex (where $q$ is
an external quark), weak boson self-energy corrections and box and pentagon
diagrams, as shown in \reffi{fig:diag_multi}.  Diagrams where the Higgs connects
to one of the external quark lines are not considered as we work in the limit of
vanishing external quark masses. Within the SM it has been found that the
leading-order contribution where a Higgs is radiated off an external bottom
quark line can give rise to a correction of about $2\%$ for a light
Higgs~\cite{07104749}. In the MSSM the Higgs coupling to bottom quarks can be
enhanced compared to the SM case. While the contributions on which we focus in
this paper can easily be supplemented by the ones where a Higgs is radiated off
an external bottom quark line, incorporation of the latter contributions would
not change our qualitative discussion below.

We incorporate the corrections to the $VVH$ vertex and the weak boson self
energy by calculating an effective $VVH$ coupling resulting from the loop and
counterterm diagrams. The most general structure of the coupling between a pair
of gauge bosons and a scalar particle is given by \cite{0403297}
\begin{eqnarray}
T^{\mu \nu} (q_{1}, q_{2}) &=& a_1(q_{1},q_{2}) g^{\mu \nu} + a_2(q_{1},q_{2})
\left[ (q_{1} q_{2}) \, g^{\mu \nu} - q_{1}^{\mu} q_{2}^{\nu} \right] +
\nonumber \\
&& a_3(q_{1},q_{2}) \epsilon^{\mu \nu \rho \sigma} q_{1 \rho} q_{2 \sigma} .
\label{eq:formfactors}
\end{eqnarray}
Here, $q_{1}$ and $q_{2}$ are the momenta of the weak bosons, and $a_{1}$,
$a_{2}$ and $a_{3}$ are Lorentz invariant formfactors.  At the tree
level, only the formfactor $a_{1}$ has a non-zero value in the SM and the MSSM:
\begin{eqnarray}
 a^{\rm SM}_{1,H^{\rm SM}WW} & = & \frac{i\, eM_{W}}{\sin\theta_{\rm W}}  
 \label{eq:a1SM} , \\
 a^{\rm MSSM}_{1,hWW} & = & \frac{i\, eM_{W}}{\sin\theta_{\rm W}}
\sin(\beta - \alpha) , \quad 
 a^{\rm MSSM}_{1,HWW}  = \frac{i\, eM_{W}}{\sin\theta_{\rm W}} \cos(\beta -
\alpha) .
\label{eq:a1MSSM}
\end{eqnarray}
At lowest order the MSSM formfactor $a_{1}$ for the lightest $\cp$-even Higgs
boson differs from the SM value of $a_{1}$ by a factor $\sin{\left(\beta -
\alpha\right)}$, which tends to $1$ in the decoupling regime, i.e.\ for $\MA \gg
\MZ$. The inclusion of higher order diagrams, however, gives rise to different
contributions to $a_{1}$ in the two models, and in general yields non-zero
values for $a_{2}$ and $a_{3}$. The approach of parametrising parts of the
one-loop contributions in terms of formfactors has the advantage of being
relatively simple, as well as being quick to calculate computationally. By
running the formfactor calculation subroutines separately, the speed of
parameter scans can be greatly enhanced, making it easier to identify
interesting regions in the supersymmetric parameter space.\footnote{Interesting
Higgs phenomenology is expected to manifest itself in these formfactors, owing
to the differences between the SM and MSSM Higgs sectors.}


There are no closed (s)fermion loops at the $qqV$ vertices.  There are, however,
counterterm diagrams, where the renormalisation constants contain self-energy
contributions from the (s)fermions.  These counterterm diagrams are separately
finite in this case.  We have calculated the corrections to the quark vertex in
two different ways. If only the contributions from the (s)fermion sector are
considered, an effective coupling can be used as there are only counterterm
contributions in this case. When considering the complete corrections, on the
other hand, we calculate the full matrix element. Finally, the box and pentagon
diagrams are included in {\tt VBFNLO} by calculating the full $2 \rightarrow 3$
matrix elements.

In order to check these procedures for internal consistency, the corrections to
the Higgs vertex have also been calculated using the full matrix elements
instead of the simpler formfactor parametrisations, and we have verified that
the two sets of results agree with each other.

\subsubsection{Higgs propagator corrections}
\label{sec:Zfac}

Higgs propagator corrections, which can be very important numerically, enter the
prediction for the mass of the external Higgs boson and are furthermore required
to ensure the correct on-shell properties of S-matrix elements involving
external Higgs bosons, i.e.\ unit residue and vanishing mixing between different
Higgs bosons on mass shell. It is convenient in this context to use finite wave
function normalisation factors, which make it easy to incorporate leading
higher-order contributions. A vertex function with an external Higgs boson $h_a$
($a = 1, 2, 3$) in general receives contributions from all three lowest-order
neutral Higgs states\footnote{In general, one also needs to consider mixing with
Goldstone bosons. For the case of weak boson fusion, however, no such
contributions occur at the one-loop level and we therefore do not consider them
here.  For completeness, contributions due to mixing with gauge bosons are
included in the calculation, although they are not significant numerically.}
according to
\begin{align}
\begin{pmatrix} \hat\Ga_{h_1} \\ \hat\Ga_{h_2} \\ \hat\Ga_{h_3} \end{pmatrix} =
\matr{\hat Z} \cdot \begin{pmatrix} \hat\Ga_h \\ \hat\Ga_H \\\hat \Ga_A
\end{pmatrix} ,
\end{align}
where the elements of the (non-unitary) matrix $\matr{\hat Z}$ have been defined
in \cite{0611326,07105320}. In the $\cp$-conserving case mixing occurs only
between the two $\cp$-even states. We calculate the wavefunction normalisation
factors and the Higgs boson masses using the program {\tt FeynHiggs}, taking
into account the full one-loop result as well as the dominant two-loop
contributions.

In our numerical discussion below we incorporate the universal wavefunction
corrections into the lowest order matrix element, so that the effect of the
genuine one-loop corrections can be discussed separately from the known
propagator-type contributions. Accordingly, in the following we use the phrase
``leading order'' for the tree-level element supplemented by the wavefunction
normalisation factors (and parametrised in terms of the loop-corrected mass of
the outgoing Higgs boson), whereas ``tree'' refers to the purely tree level
diagrams without the wavefunction normalisation factors (parametrised also in
this case in terms of the loop-corrected Higgs mass).

In principle there is a choice between treating the universal wavefunction
corrections as an ``additive'' correction, i.e.\ absorbing them into the
tree-level part of the amplitude only, or as a ``multiplicative'' correction,
i.e.\ applying them both to the tree-level and the one-loop part of the
amplitude. The difference between the two options is of higher order. For the
results shown below, in which the wavefunction corrections are treated as
``additive'', the numerical difference between the two options is insignificant.
Sizable effects are possible, however, in ``extreme'' regions of the parameter
space, for instance in the non-decoupling regime of $\cp$-violating scenarios.
In cases like this the inclusion of the wavefunction corrections in the loop
part of the amplitude can have an impact on the shape of azimuthal angle
distributions. As the universal wavefunction corrections are not the main focus
of the present paper, we will not discuss this issue any further here.

\subsubsection{Real corrections}
\label{sec:realcorr}

As mentioned above, we work in the limit of vanishing quark mass for the
external (1st and 2nd generation) quarks. We regularise the IR and collinear
divergences by a small photon mass and small quark masses, respectively, and use
the dipole subtraction formalism as described in \cite{9904440}. As an
additional check on the IR finiteness of the results, we have also implemented
the soft photon approximation (see e.g.\ \cite{Denner}) as an alternative to
dipole subtraction.  Matrix elements for processes with real photon emission
have been calculated using helicity amplitudes \cite{313npb560}, and have
been numerically compared with matrix elements generated with {\tt Madgraph}
\cite{MadGraph} for individual phase space points.

\subsection{Renormalisation}
\label{renorm}

We perform the renormalisation of the parameters and fields as outlined in
\cite{0611326}. While the algebraic structure of the counterterms for the $qqV$
vertex and $VV$ self energy are the same in the MSSM as in the SM, the
counterterms for the $VVH$ vertices contain contributions from the
renormalisation of $\tb$ and off-diagonal contributions from the Higgs field
renormalisations. For the $\cp$-conserving case the explicit form of these
counterterms has been given in \cite{0211204}. In the $\cp$-violating case there
are non-zero counterterm contributions for all three vertices of the kind
$VVh_a$, $a = 1, 2, 3$. The relevant MSSM counterterms were implemented into a
{\tt FeynArts} model file. The implementation of our results into {\tt VBFNLO}
has three options for parametrising the electromagnetic coupling in the Born
level cross section.  In the code the electromagnetic coupling can be
parameterised by $\alpha(0)$, $\alpha(\MZ)$ and via the Fermi constant, $G_F$.
In the latter case the relation
\begin{equation}
 \alpha \equiv \alpha(0)
 = \frac {\sqrt{2} G_{F} M_{W}^{2}} {\pi \left(1 + \Delta r\right)}
\left( 1 - \frac {M_{W}^{2}} {M_{Z}^{2}} \right)
\label{eqn:alpha}
\end{equation}
is employed, where the quantity $\Delta r$ contains higher-order corrections to
muon decay, see \cite{0604147,07102972}. The charge renormalisation counterterm
is adjusted according to the chosen option.

We have checked that the UV divergences cancel not only for the full result but
also separately for the (s)top / (s)bottom and for the (s)fermion contributions.
Furthermore we have verified for the (s)fermion contributions that the loop
corrections, plus the appropriate counterterms, from the $qqV$ vertex, the weak boson self energy and the $VVH$
vertex are all separately finite. We have also checked, algebraically and
numerically, that the weak boson field renormalisation constants drop out in the
sum of all counterterm contributions. The parameters used for regularising the
UV divergence and the IR divergencies (quark mass and photon mass) can all be
varied numerically in the code.  We have verified that our result has no
dependence on any of these parameters.


\section{Numerical results}
\label{sec:num_res}

Unless otherwise stated, we use the PDF set MRST2004qed \cite{0411040}, as this
includes $\mathcal{O}(\alpha)$ QED corrections (thus allowing photon induced
processes to be considered).  
 There is no LO PDF set associated with
MRST2004qed, and so we use these distributions at both NLO and LO\footnote{This
also facilitated comparisons with Ref.~\cite{07104749}, where the same PDF set was
used.}.  The gauge coupling is parametrised by $G_{F}$,%
\footnote{One might argue that it would be preferable to parametrise the
result in terms of $\alpha(0)$, since this choice is used in the PDF
set. However, as the parametrisation in terms of $G_{F}$ is known to
absorb numerically relevant electroweak loop corrections we regard the
latter parametrisation as the more appropriate one for this process. 
For comparison, we have also calculated the results using the
parametrisation in terms of $\alpha(0)$. While, as expected, the
relative corrections are strongly affected by the choice of
parametrisation, the difference between the NLO cross sections using the
two parametrisations (which is formally a two-loop effect) amounts to a
maximum of 1\% for small Higgs masses and decreases with increasing
Higgs mass.}
and a centre of mass
energy of 14 TeV is used.  We normally set $m_{t} = 172.6$ GeV~\cite{08031683},
and for the other parameters we use the values given by the Particle Data
Group~\cite{PDB}.  By default, we use $M_{W}$ as both the renormalisation and
factorisation scale.

\subsection{Cuts and non-WBF processes}
\label{sec:cuts}

By default, the cuts used here are those described in \cite{07104749}:
\begin{eqnarray}
 p_{T_{j}} &>& 20 \mbox{ GeV} \nonumber \\
 \mid y_{j} \mid &\geq& 4.5  \nonumber \\
 \Delta y_{ij} \equiv \mid y_{j_{1}} - y_{j_{2}} \mid &>& 4 \nonumber \\
 y_{j_{1}} \cdot y_{j_{2}} &<& 0,
\end{eqnarray}
where $p_{T_{j}}$ is the transverse momentum of a jet, and $y_{j}$ is its
rapidity.  In addition, the $k_{T}$ algorithm is used to reconstruct jets from
the final state partons, using the parameters
\begin{eqnarray}
 R_{jj} &\geq& 0.8 \nonumber \\
 \mid \eta \mid &<& 5 ,
\end{eqnarray}
where $R_{jj}$ is the $R$ separation of the two jets, and $\eta$ is the
pseudorapidity of the partons.  These cuts ensure that the signal is relatively
clean and that the effect of processes such as the s-channel Higgsstrahlung
process is small \cite{0611281,07093513,07104749}. Consequently, the
Higgsstrahlung process is not included in this work.

\subsection{Comparison with the literature}
\label{sec:comp}

As a first step, the leading order result in the SM was checked against the
result obtained using \tt MadGraph \rm \cite{MadGraph}. The results were found
to agree to within the numerical accuracy of the respective codes.  
Additionally, the tree-level matrix elements for Higgs production via weak boson
fusion plus a photon were also compared with \cite{1006.4237}, and found to be
in full agreement.

\begin{table}
\caption{\small{Comparison of our results, as implemented into the code {\tt
VBFNLO} (labelled ``this work''), for the leading order (LO) cross section and
the full one-loop contribution (NLO, containing both QCD and electroweak
corrections) to the weak boson fusion (t) channel with those obtained using
\tt{HAWK}\rm, the code developed in \citere{07104749}.}
\label{tab:wbfMC_comp}}
\begin{center}
\begin{tabular}{lccc}
\hline\noalign{\smallskip}
$M_{H}$ [GeV] & 120 & 150 & 200 \\
\noalign{\smallskip} \hline\noalign{\smallskip}\hline
$\sigma_{LO}$, \tt{HAWK} \rm [fb] & 1876.96 $\pm$ 1.59 & 1589.87 $\pm$ 1.25 &
1221.40 $\pm$ 0.87 \\
$\sigma_{LO}$, this work [fb] & 1876.66 $\pm$ 1.32 &
1590.19 $\pm$ 1.10 & 1221.26 $\pm$ 0.82 \\
\noalign{\smallskip}  \hline\noalign{\smallskip}
$\sigma_{NLO}$, \tt{HAWK} \rm [fb] & 1637.85 $\pm$ 3.22 & 1387.36 $\pm$ 2.40 &
1074.14 $\pm$ 1.72 \\
$\sigma_{NLO}$, this work [fb] & 1634.54 $\pm$ 2.39 & 1387.37 $\pm$ 2.09 &
1073.08 $\pm$ 1.54 \\
\noalign{\smallskip}\hline
\end{tabular}
\end{center}
\end{table}

We next compare the complete one-loop result for the weak boson fusion channel
in the SM with the result obtained in \citere{07104749}. Accordingly, we compare
our result with the t-channel contribution given by the code \tt{HAWK}\rm, which
was developed in \citere{07104749}.\footnote{Contributions from the s-channel
and t/u interference are numerically small, below the level of $\sim$0.5\%, once
weak boson fusion cuts have been applied. We furthermore have not included
photon induced processes in this comparison, as in general their s- and
t-channel contributions are not separately gauge-invariant, but we have verified
that our results for the photon induced processes are in good agreement with
\citere{07104749}.} Table~\ref{tab:wbfMC_comp} shows a comparison for on-shell
Higgs production between {\tt VBFNLO}, incorporating our results, and the result
obtained using \tt{HAWK}\rm, with all parameters and cuts set to match
\citere{07104749}. Table~\ref{tab:wbfMC_comp} shows that both the leading order
results of the two codes as well as the predictions for the cross section
including the complete QCD and electroweak one-loop corrections in the SM fully
agree with each other within the numerical uncertainties. 

We now turn to the comparison with the results for the purely supersymmetric
corrections to weak boson fusion in the MSSM with real parameters given in
\citere{Michael}. The separation into ``pure SUSY'' and ``SM-type''
contributions can easily be performed as long as one only considers loop
contributions from SM fermions and their scalar superpartners. Going beyond the
(s)fermion contributions, however, this distinction is less obvious owing to the
increased complexity of the Higgs sector in the MSSM as compared to the SM case.
The authors of \citere{Michael} have chosen to define the ``pure SUSY''
corrections for the production of the light $\cp$-even Higgs boson according to
\begin{eqnarray}
 \sigma_{SUSY} = \sigma_{MSSM} - \sin^{2}(\beta - \alpha) \, \sigma_{SM} ,
 \label{eqn:pureSUSY}
\end{eqnarray}
which ensures that $\sigma_{SUSY}$ contains only IR-finite virtual contributions
(the factor $\sin^{2}(\beta-\alpha)$ is the ratio of the squared lowest order
coupling of the light $\cp$-even Higgs to two weak bosons over the corresponding
coupling of a SM Higgs, see \refeqs{eq:a1SM}, (\ref{eq:a1MSSM})). In comparing
with \citere{Michael} we focus on the supersymmetric contributions to the $VVh$
vertex.\footnote{Here, as in Ref. \cite{Michael}, we include contributions from
photon fusion and photon-Z fusion.} Table~\ref{tab:Michael_comp} shows a
comparison of our results with the ones of \citere{Michael} for the relative
impact of the pure SUSY loop corrections, defined according to
\refeq{eqn:pureSUSY}, on the total Higgs production cross section. In order to
enable a comparison with the relative corrections given in \citere{Michael} we
have expressed the Higgs propagator corrections (see \refse{sec:Zfac}) as part
of the loop contributions rather than absorbing them into the leading order
result as we do elsewhere in this paper.\footnote{It turns out that our result
for the leading order cross section differs from the value stated in
\citere{Michael}.}

\begin{table}
\caption{\small{Comparison of percentage corrections to the total Higgs
production cross section arising from contributions to the $VVh$ vertex of pure
SUSY type, defined according to \refeq{eqn:pureSUSY}, with the results presented
in \citere{Michael}.  The column labelled ``This work'' gives our results,
incorporating Higgs propagator corrections up to the two-loop level. The column
labelled ``Tuned result'' (with Monte Carlo errors) was obtained by adapting our
calculation to the prescriptions used in \citere{Michael} (see text). The column
labelled ``Propagator-type corrections'' gives the percentage correction arising
from the universal wavefunction normalisation factors, see \refse{sec:Zfac}. The
right-most column shows the results as given in \citere{Michael}.}
\label{tab:Michael_comp} }
\begin{center}
\begin{tabular}{ccccc}
\hline\noalign{\smallskip}
SPS & This work & Tuned result & Propagator-type corrections  & Ref.\ \cite{Michael} \\
\noalign{\smallskip} \hline\noalign{\smallskip}\hline
1a & -0.210 & -0.365 $\pm$ 0.072 & 3.231 & -0.329 \\
1b & 0.044 & -0.204 $\pm$ 0.069 & 3.431 & -0.162\\
2 & -0.046 & -0.224 $\pm$ 0.067 & 3.539 & -0.147\\
3 & -0.028 & -0.214 $\pm$ 0.068 & 3.557 & -0.146\\
4 & -0.065 & -0.274 $\pm$ 0.063 & 3.173 & -0.258\\
5 & -0.651 & -0.619 $\pm$ 0.058 & 1.970 & -0.606\\
6 & -0.108 & -0.281 $\pm$ 0.063 & 3.395 & -0.226\\
7 & -0.052 & -0.246 $\pm$ 0.065 & 3.691 & -0.206\\
8 & -0.007 & -0.216 $\pm$ 0.067 & 3.766 & -0.157\\
9 & 0.031 & -0.190 $\pm$ 0.071 & 3.956 & -0.094\\
\noalign{\smallskip}\hline
\end{tabular}
\end{center}
\end{table}
There are several differences between our approach and that used in
\citere{Michael}, related in particular to the treatment of higher-order
corrections in the Higgs sector. We use the tree level Higgs masses and mixing
angle for all Higgs bosons occuring within loop diagrams,\footnote{This
ensures
the UV finiteness of the Higgs self energies.  For the investigated
parameter range, the effect of using corrected Higgs masses within the loops is
small, leading to a difference of up to 0.4\% in the NLO cross sections for the
light Higgs, and below 1\% for the heavy Higgs in phenomenologically interesting 
regions of parameter space.} whereas \citere{Michael} uses
loop-corrected masses and couplings.  In our work, we incorporate contributions
up to the two-loop order in the Higgs propagator-type
corrections\footnote{For the SPS points investigated here, the effect of the
two-loop contributions is very small -- at the per-mille level -- but their
effect can of course be larger in the non-decoupling region.} entering the
predictions for the Higgs masses and the Higgs wavefunction normalisation
factors (see \refse{sec:Zfac}), while in \citere{Michael} the contributions to
the Higgs field renormalisations are restricted to the one-loop level. 
We have added a column labelled
``Tuned result'' in Table~\ref{tab:Michael_comp} that has been obtained using a
specially tuned version of our code, where the treatment of the higher-order
corrections in the Higgs sector has been performed in accordance with the
prescription in \citere{Michael}.\footnote{Note that slightly different
versions of \tt FeynHiggs \rm were used, leading to small differences in the
values of the Higgs parameters.}

The comparison in Table~\ref{tab:Michael_comp} has been carried out for the SPS
benchmark points~\cite{SPS}, where the same low-energy input parameters have
been used as in \citere{Michael}. Furthermore, the electromagnetic coupling
constant is set to $\alpha(0)$, the PDF set MRST2002nlo~\cite{MRST2002} is
applied, the top mass is set to $m_{t} = 170.9$~GeV, the renormalisation and
factorisation scale is set to $M_{h}$, and the same set of cuts is used as in
\citere{Michael}. For illustration, in the column labelled ``Propagator-type
corrections'' in Table~\ref{tab:Michael_comp} we separately show the percentage
loop correction arising from the universal wave function normalisation factors
(see \refse{sec:Zfac}).

The relative corrections for the different SPS benchmark points shown in
Table~\ref{tab:Michael_comp} are found to be rather small, well below the level
of 1\%. This turns out to be a consequence of cancellations between the
universal propagator-type corrections, which are at the level of 3--4\%, as seen
in the fourth column of Table~\ref{tab:Michael_comp}, and the process-specific
genuine vertex corrections.\footnote{Note that the propagator corrections used
in this comparison are larger than those used in the rest of this paper, owing
to a different scale choice.} The latter tend to overcompensate the positive
correction arising from the propagator-type contributions, yielding overall a
negative correction at the sub-percent level for most of the SPS points. It
should furthermore be noted in this context that all SPS points belong to the
decoupling region of the supersymmetric parameter space, where the couplings of
the light $\cp$-even Higgs are SM-like, i.e.\ no large SUSY loop effects on the
Higgs couplings are expected in this parameter region. The comparison between
our results (labelled as ``This work'') and the ones quoted in \citere{Michael}
shows reasonably good agreement, with absolute deviations at the level of 0.2\%
or below. The agreement further improves if our ``Tuned result'' (where the
treatment of the higher-order corrections in the Higgs sector has been performed
in accordance to the prescription in \citere{Michael}, as explained above) is
used for the comparison. The remaining deviations between the ``Tuned result''
and the results of \cite{Michael} arise from a combination of small
factors, such as slight remaining differences in the calculation of the Higgs
sector, the differences in the procedures used to implement the cuts, and the
numerical inaccuracy
inherent in the Monte Carlo integration.  These Monte Carlo errors are given in the column ``Tuned Result'' of Table~\ref{tab:Michael_comp}, and are obtained by adding the errors on the LO and NLO cross section in quadrature.  As can be seen, once these numerical errors are taken into account, the tuned results agree with those presented in Ref.~\cite{Michael}\footnote{Owing to the very small ``pure SUSY'' corrections found using our tuned procedure at these particular parameter points, the Monte Carlo error seems large in comparison: the errors quoted on the percentage corrections correspond to errors below the per-mille level on the cross sections themselves.}.

\subsection{Total cross sections and distributions}
\label{sec: results}

\begin{figure}[htb!]
     \subfigure[Total cross section for Higgs production as a function of
$M_{H}$ in the Standard Model.]{
         \label{fig:SM_sigma}
         \resizebox{0.46\hsize}{!}{\includegraphics*{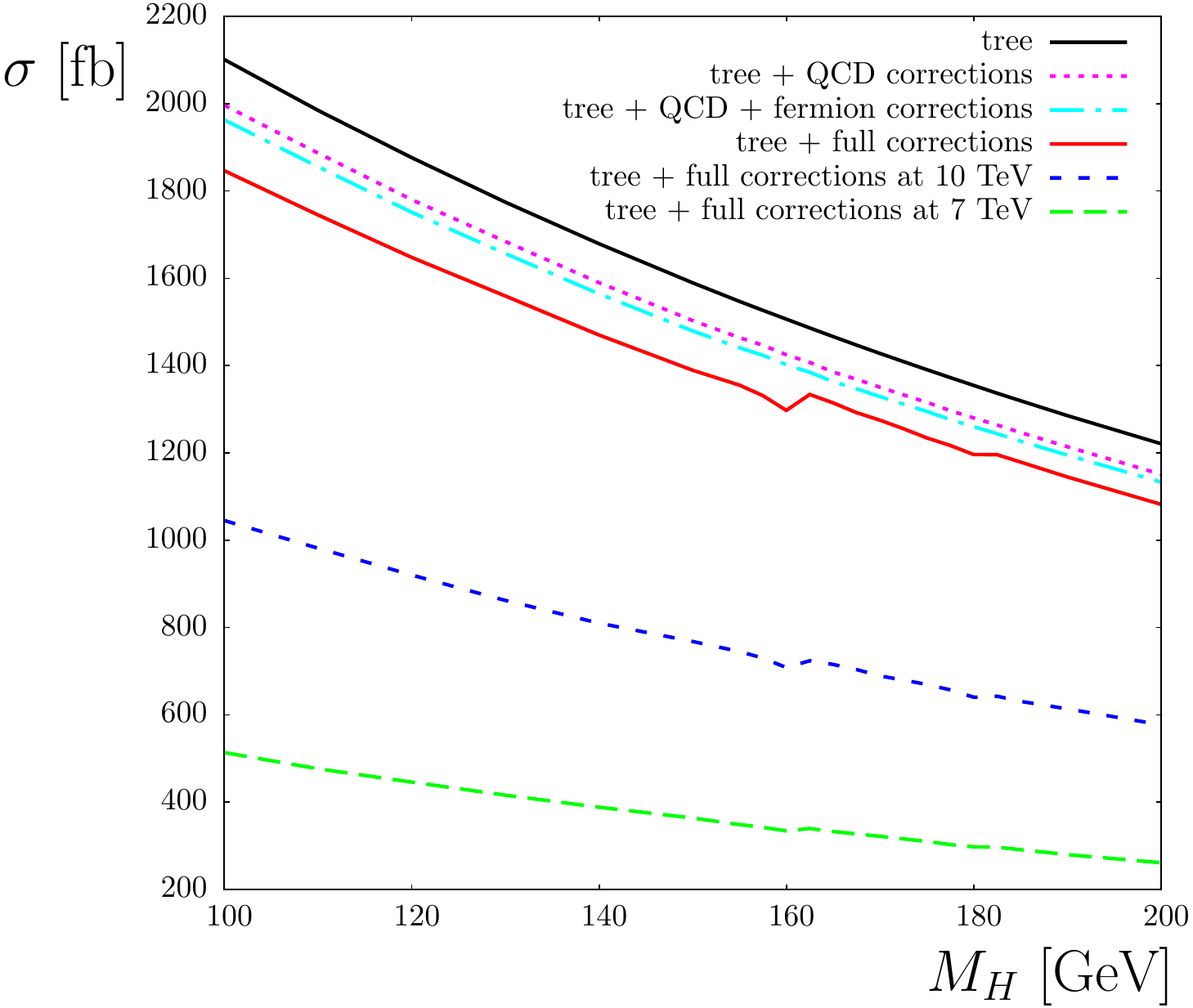}}
         }
     \hspace{.3cm}
     \subfigure[Percentage higher order corrections for Higgs production as a
function of $M_{H}$ in the Standard Model.]{
         \label{fig:SM_sigmaF}
         \resizebox{0.47\hsize}{!}{\includegraphics*{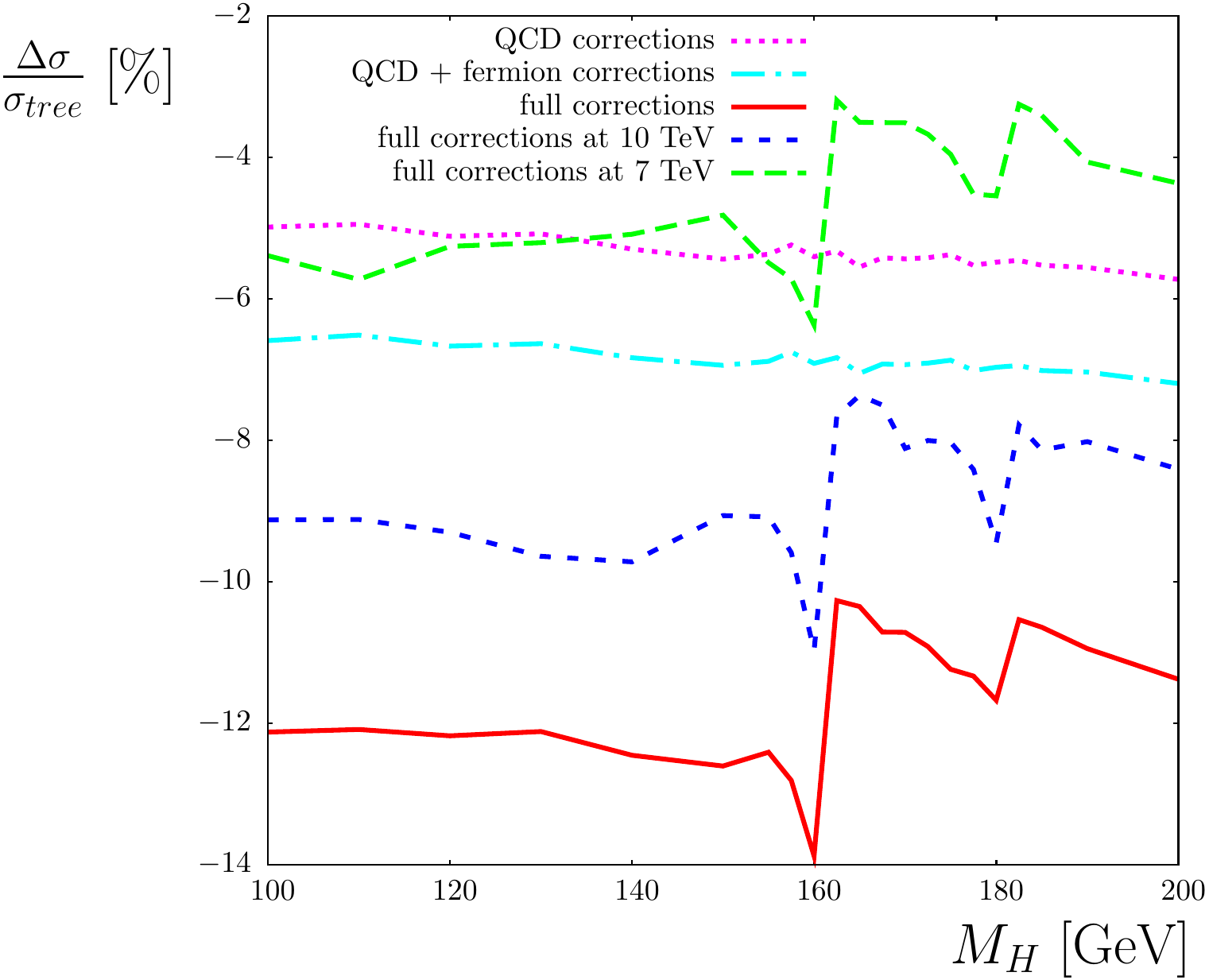}}
         }  
   \vspace{-0.5cm}
     \caption{Results for Higgs boson production via weak boson fusion 
in the Standard Model. The full one-loop result, labelled 
``tree $+$ full corrections'' is compared with various approximations (see
text).}
     \label{fig:SM_results}
\end{figure}

Figure \ref{fig:SM_sigma} shows the total cross section in the Standard Model
for a range of $M_{H}$, obtained from {\tt VBFNLO} incorporating our results.
The curve labelled ``tree $+$ full corrections'' shows the full one-loop result
in the SM for Higgs production in weak boson fusion at the LHC with 14 TeV,
using the input values and cuts as specified above. The full one-loop result is
compared with the tree-level result (labelled ``tree''), the result incoporating
only QCD corrections (``tree $+$ QCD corrections'') and the result incorporating
in addition fermion-loop corrections (``tree $+$ QCD $+$ fermion corrections'').
For illustration the full result is also shown for energies of 10 TeV and 7 TeV,
corresponding to a reduction of the cross section by a factor of approximately
1.8 and 3.8 respectively (for a Higgs mass of $120 \gev$ compared to the cross
section at 14 TeV). Figure \ref{fig:SM_sigmaF} shows the percentage corrections.
One can see from the plot that the QCD and the electroweak corrections are of
similar size, being of order 5\%, and enter with the same sign. It is
interesting to note that the non-fermion contributions to the loop corrections
are significant, causing a further reduction in the cross section.  In our
parametrisation of the result, the (bosonic) box- and pentagon-type
contributions turn out to be numerically small.  In Figure \ref{fig:SM_sigmaF},
the thresholds at $M_{H} = 2 M_{W}$ and $M_{H} = 2 M_{Z}$ are clearly visible. 
The reduction in the percentage corrections for lower centre of mass energies
originates primarily in the QCD corrections.

\begin{figure}[htb!]
\begin{center}
          \resizebox{0.46\hsize}{!}{\includegraphics*{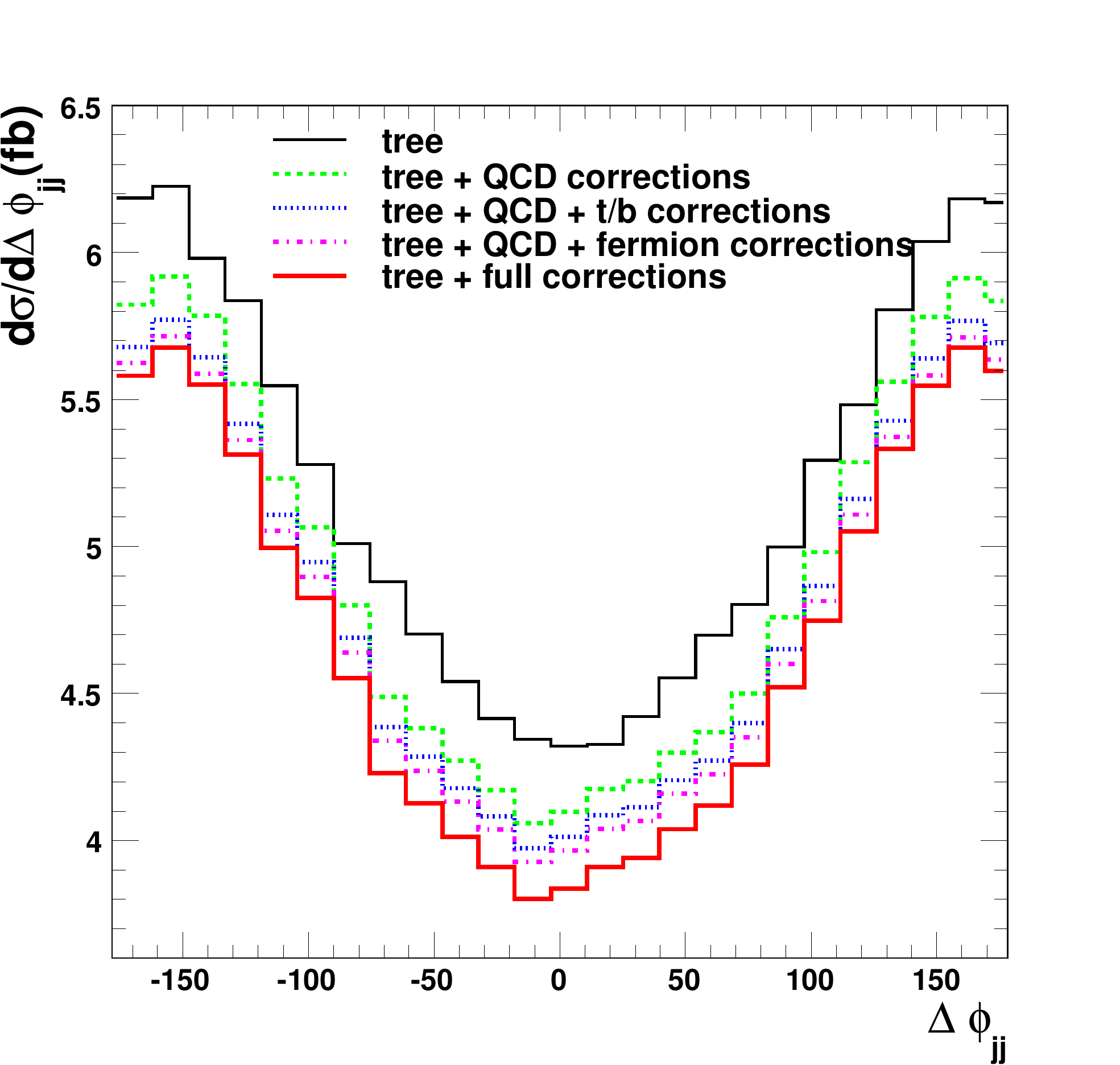}}
\end{center}
  \vspace{-1cm}
     \caption{Azimuthal angle distribution in the Standard Model, with $M_{H}
=$ 120 GeV.  The full one-loop result, labelled 
``tree $+$ full corrections'' is compared with various approximations (see
text).}
\label{fig:SM_phi}
\end{figure}

As an example for a differential distribution we show the azimuthal angle
distribution in Figure~\ref{fig:SM_phi}. This distribution is of particular
interest in determining the structure of the coupling between the Higgs boson
and the weak boson pairs~\cite{0609075}, since its shape is in principle
sensitive to the relative values of the formfactors $a_{1}$, $a_{2}$ and
$a_{3}$, as defined in \refeq{eq:formfactors}. The distribution is shown for a
mass $M_{H} = 120 \gev$ in the SM. The relative impact of the different types of
corrections is as in Figure~\ref{fig:SM_sigma} (here, we also present the result
containing the contributions from the third generation quarks in addition to the
QCD corrections (``tree $+$ QCD $+$ t/b corrections''), which shows that the
fermion loop contributions are dominated by the third generation quarks). The
shape of the distribution turns out to be affected only mildly by the
higher-order corrections. This can be understood from the fact that only the
formfactor $a_{1}$ receives significant corrections in the SM, at the level of
1--2\%, while the formfactors $a_{2}$ and $a_{3}$ in the SM are extremely small
(approximately 5 and 10 orders of magnitude smaller than $\Delta a_{1}$
respectively) so that the corresponding effects will not be experimentally
detectable at the LHC~\cite{Markus}.

\begin{figure}[htb!]
     \subfigure[Loop corrections $\Delta a_{1}$ to the formfactor $a_{1}$ as a 
percentage of the tree level, $a_{1}^{\rm LO}$, for benchmarks in the MSSM with 
real parameters.]{
         \label{fig:mssm_Wa1}
          \resizebox{0.45\hsize}{!}{\includegraphics*{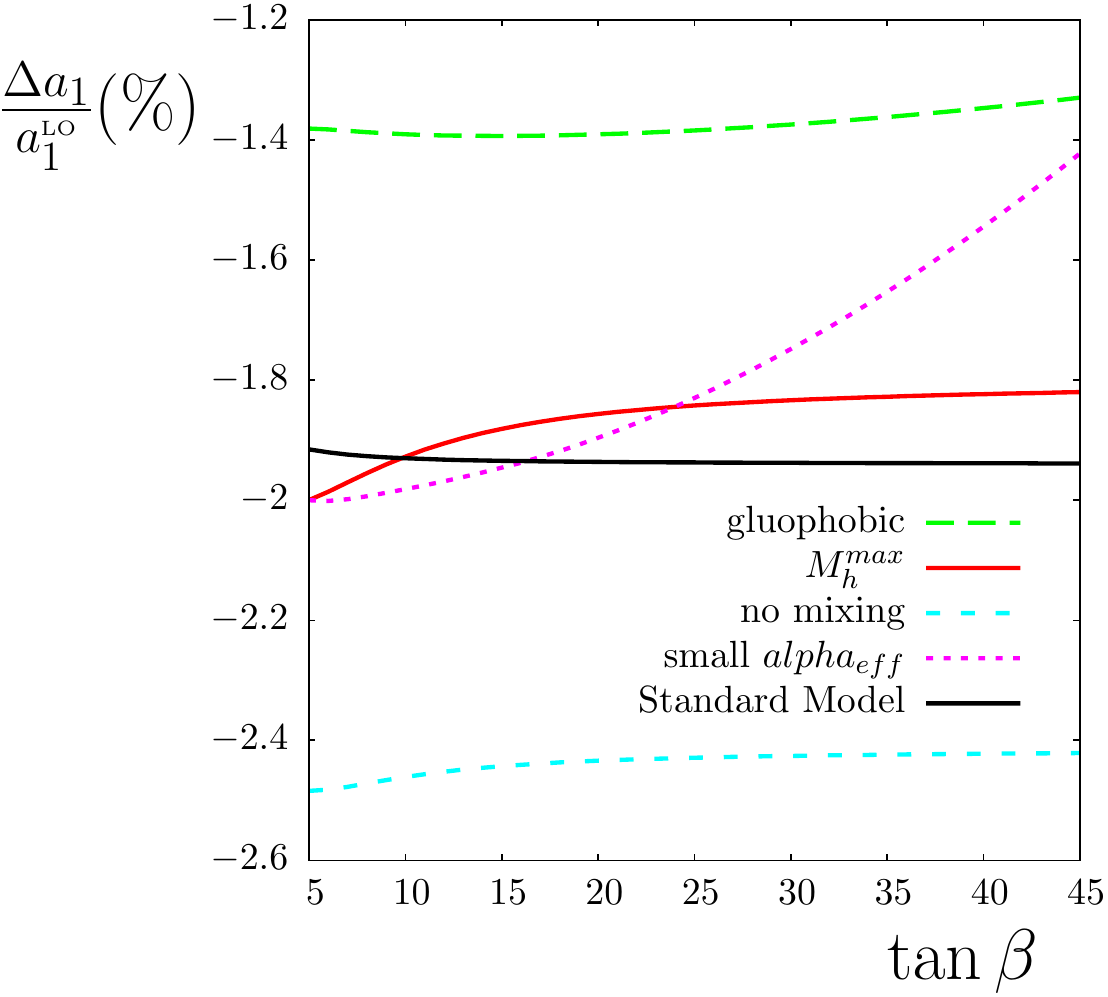}}
         }
     \hspace{.3cm}
     \subfigure[Corrected value of the formfactor $a_{1}$ for all benchmarks.]{
         \label{fig:mssm_Wa1C}
          \resizebox{0.4\hsize}{!}{\includegraphics*{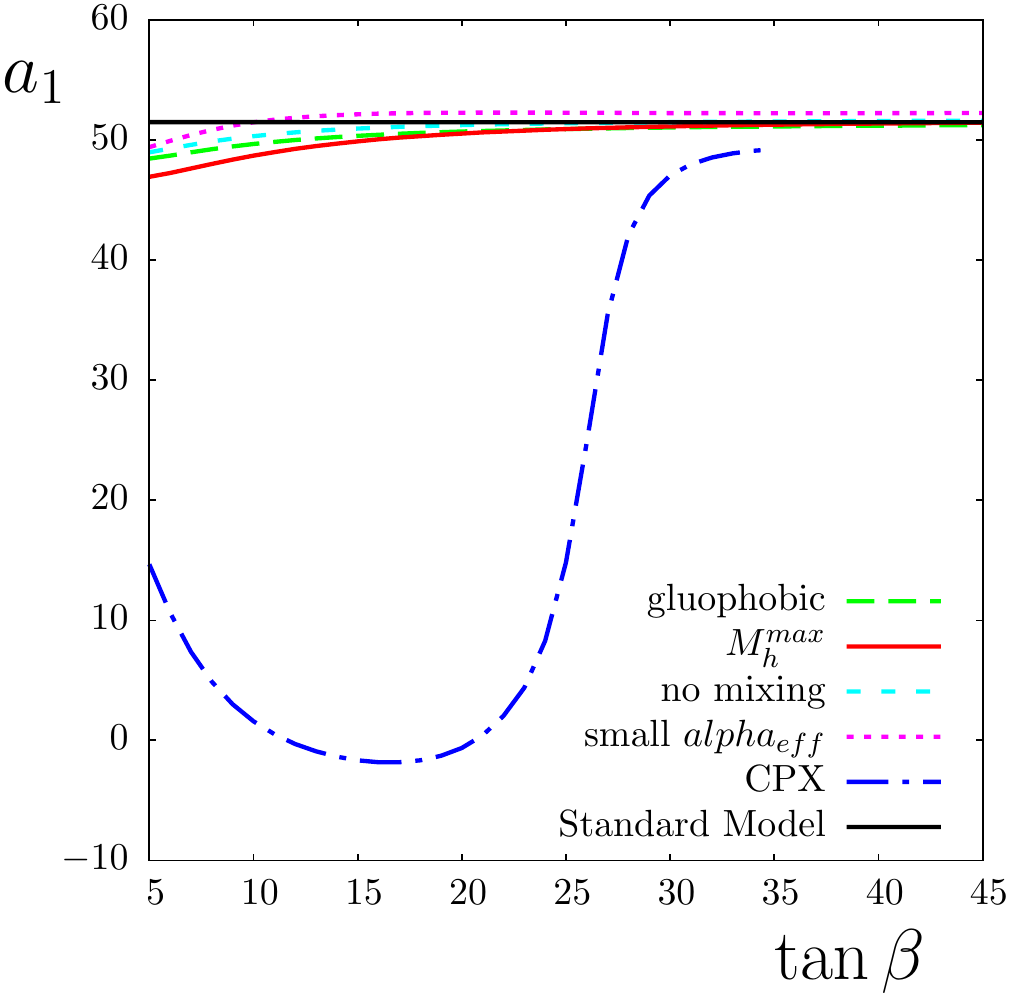}}
         }
     \hspace{.3cm}
     \subfigure[Value of the formfactor $a_{2}$.]{
         \label{fig:mssm_Wa2}
          \resizebox{0.45\hsize}{!}{\includegraphics*{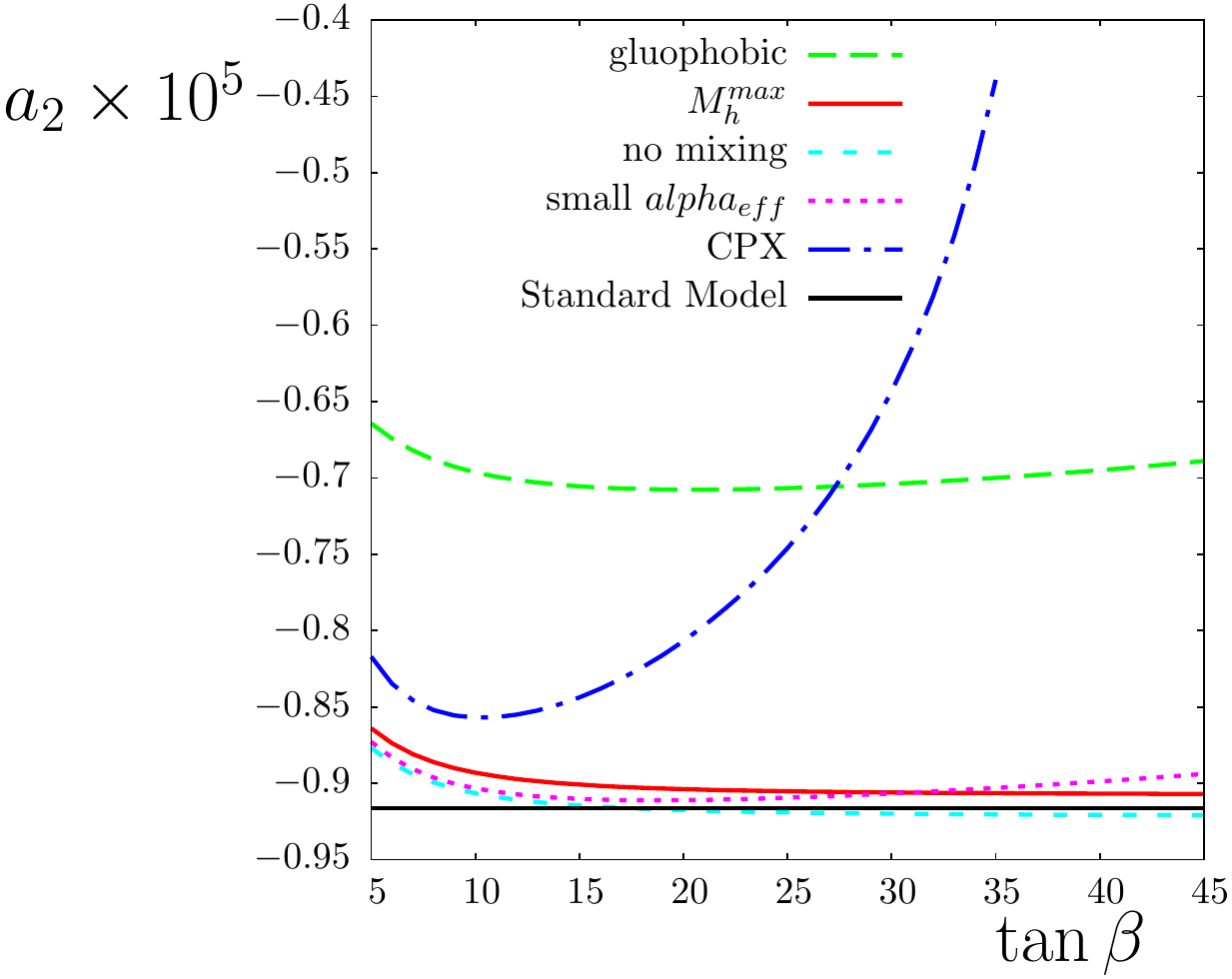}}
          }
     \hspace{.3cm}
     \subfigure[Value of the formfactor $a_{3}$.]{
         \label{fig:mssm_Wa3}
          \resizebox{0.45\hsize}{!}{\includegraphics*{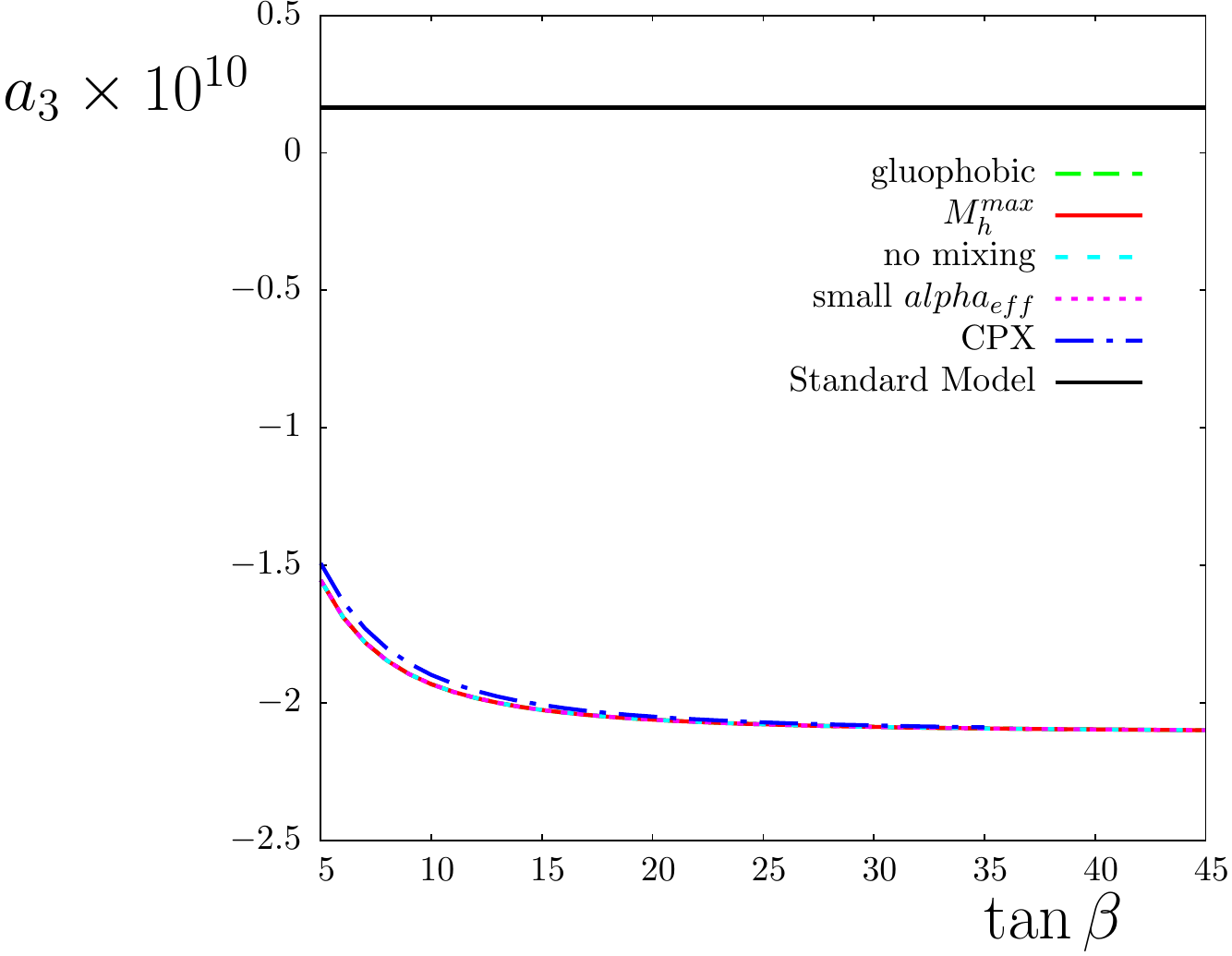}}
          }
  \vspace{-0.5cm}
     \caption{Corrections from fermion and sfermion loops to the 
formfactors of the $WWh_{1}$ vertex in the MSSM as
a function of $\tan\beta$, with $\MA = 150\gev$ (for the CPX scenario,
$\MHpm = 150\gev$ is used).  For comparison, the SM
formfactors are also shown, with a Higgs mass that matches the light
$\cp$-even Higgs mass in the \Mhmax\ scenario.}
     \label{fig:mssm_ff}
\end{figure}

Moving to the case of the MSSM, we first investigate the impact of the loop
corrections from fermions and sfermions on the formfactors $a_1$, $a_2$, $a_3$
for the $WWh$ vertex. This is shown in \reffi{fig:mssm_ff}, where the MSSM
predictions in different benchmark scenarios are given as a function of $\tb$
for $\MA = 150\gev$. The ($\cp$-conserving) \Mhmax, no-mixing, small $\aeff$ and
gluophobic benchmark scenarios have been defined in \citere{0202167}, while for
the ($\cp$-violating) CPX scenario we use the definition given in
\citere{ex0602042}, except that for the trilinear coupling parameter $A_{t}$ we
use the (on-shell) value of $900\gev$ (for the CPX scenario we use $\MHpm =
150\gev$ rather than $\MA = 150\gev$). For comparison, the result in the SM is
also shown, where the value of the Higgs mass has been set to the value obtained
for the light $\cp$-even Higgs mass in the \Mhmax\ scenario. 

\reffi{fig:mssm_Wa1} illustrates that the corrections to $a_{1}$ can be larger
in the $\cp$-conserving benchmarks than in the SM case, but are still typically
of the order of a few per cent (note that, unlike in \refse{sec:comp}, these
results include all fermion and sfermion diagrams involved in the corrections to
the formfactors, rather than the purely supersymmetric corrections). The
situation is different in the CPX scenario, see \reffi{fig:mssm_Wa1C}, where
$a_{1}$ has an exceedingly small value in certain regions of parameter space,
due to the loop-induced mixing between the three neutral Higgs bosons. As in the
SM case, the contributions to $a_{2}$ and $a_{3}$ (see \reffi{fig:mssm_Wa2},
\ref{fig:mssm_Wa3}) turn out to be very small.\footnote{Note that, as
expected, the sfermions do not contribute to the value of $a_{3}$.  The
behaviour of $a_{3}$ as a function of $\tan \beta$ is purely the result of
different couplings to the Higgs boson.} 

\begin{figure}
     \subfigure[Total cross section.]{
           \resizebox{0.46\hsize}{!}{\includegraphics*{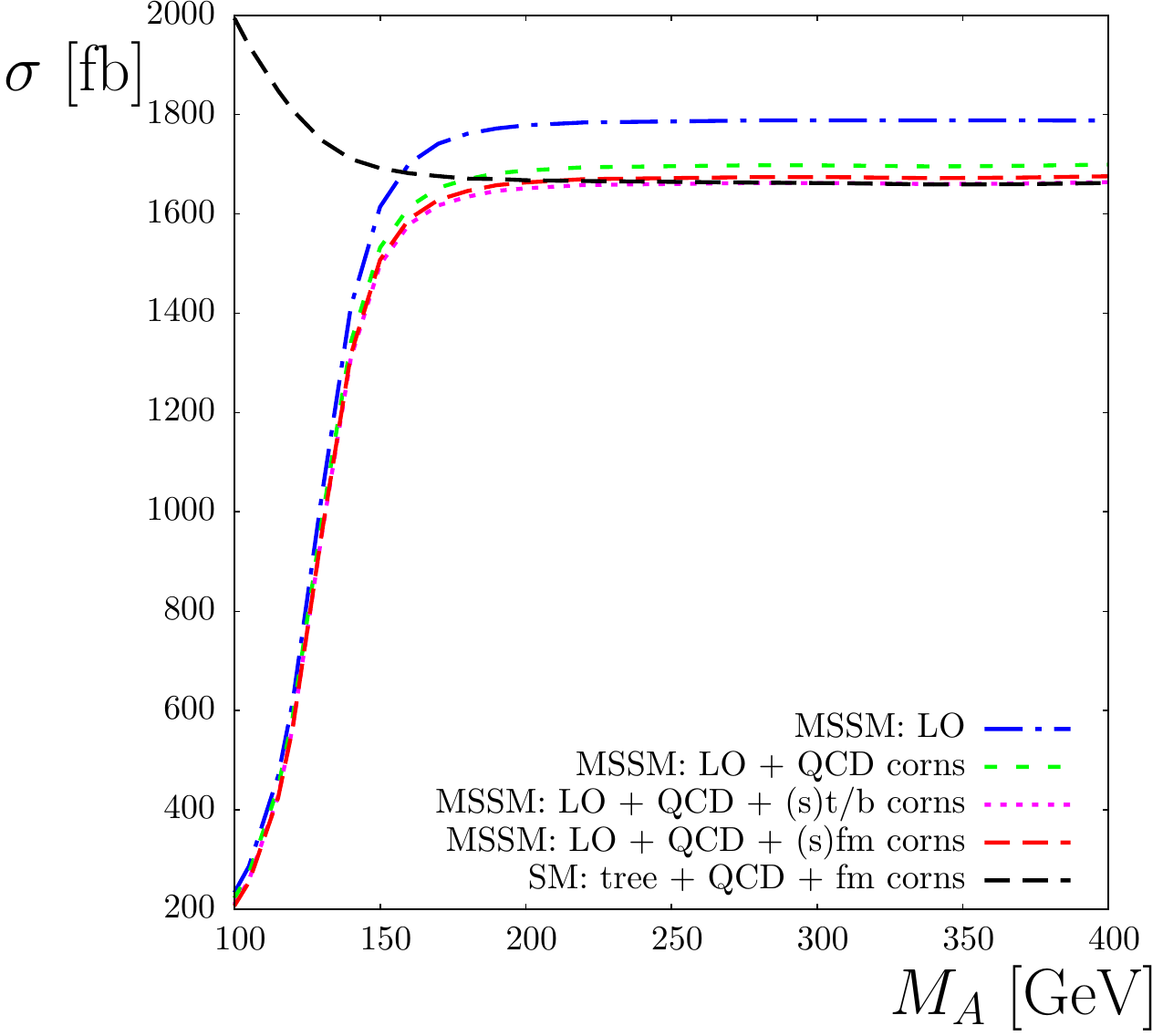}}
         }
     \hspace{.3cm}
     \subfigure[Percentage corrections.]{
          \resizebox{0.47\hsize}{!}{\includegraphics*{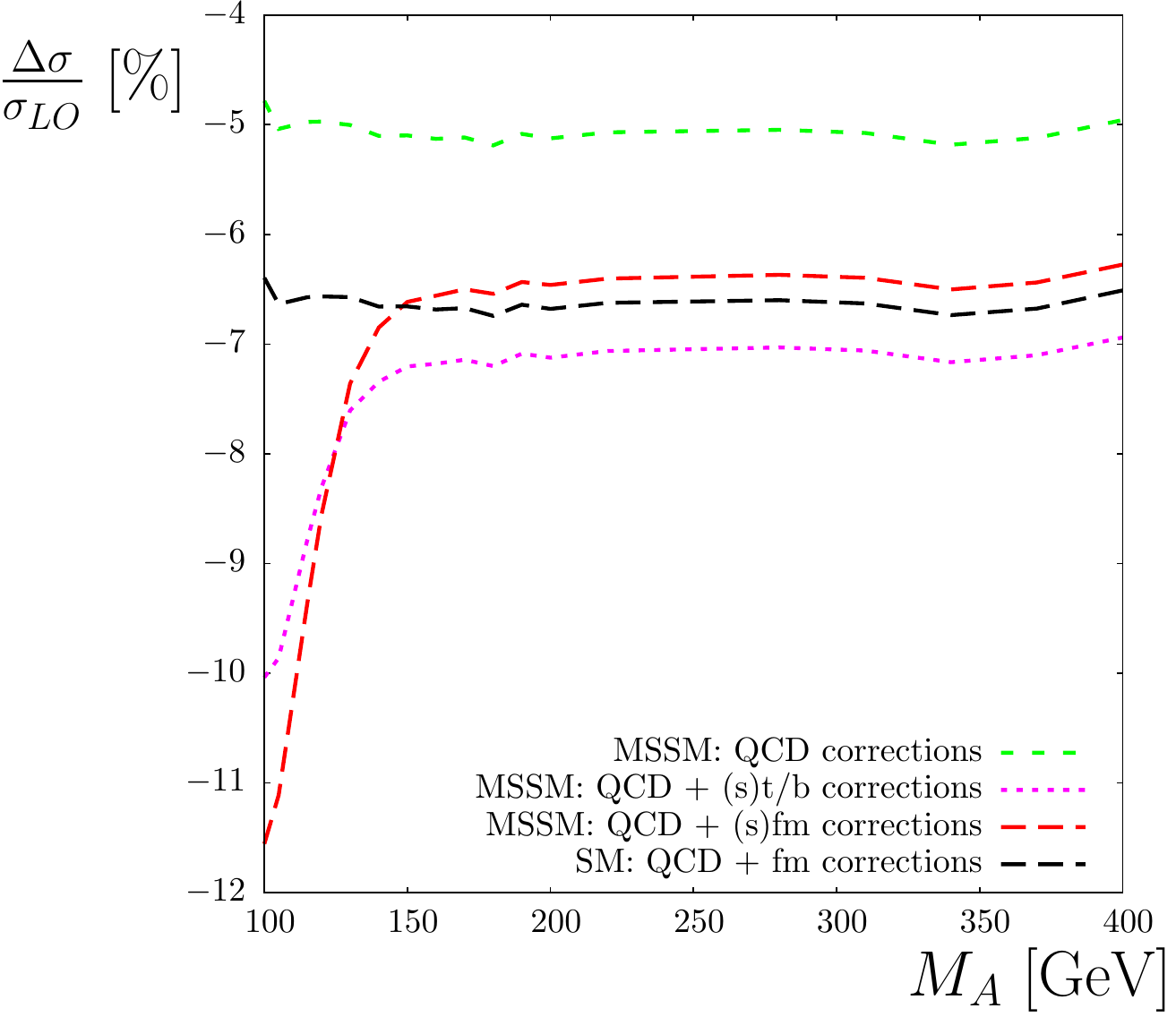}}
         }
  \vspace{-0.5cm}
     \caption{Light Higgs boson $h$ production as a function of $\MA$ in the
\Mhmax\
scenario, with tan$\beta$ = 10.}
     \label{fig:mhm_sigmah0}
\end{figure}

We next consider the total cross section for the production of the light MSSM
Higgs boson $h$ in the \Mhmax\ scenario as a function of $\MA$. We begin in
\reffi{fig:mhm_sigmah0} by comparing the leading order (LO) cross section in the
MSSM (as explained above, the LO cross section contains the effect of the
universal wavefunction normalisation factors and is evaluated at the
loop-corrected value of the Higgs mass) with the prediction where SM QCD
corrections are included (labelled ``LO $+$ QCD corrections'') and with the
predictions where in addition the loop corrections from the third generation
quarks and their scalar superpartners (``LO $+$ QCD $+$ (s)t/b corrections'')
and from all three generations of quarks and leptons and their scalar
superpartners (``LO $+$ QCD $+$ (s)fermion corrections'') are included. The
range in $\MA$ displayed in \reffi{fig:mhm_sigmah0} represents a variation in
the mass of the lightest Higgs boson from $\approx 98\gev$ to $\approx 130\gev$.
For comparison, the Standard Model cross section for the corresponding value of
the Higgs mass is also shown. \reffi{fig:mhm_sigmah0}(b) shows the relative size
of the higher-order corrections, normalised to the leading-order prediction (we
parametrise the leading order cross section in terms of the Fermi constant
$G_{F}$, see \refeq{eqn:alpha}, taking into account the appropriate
contributions to $\Delta r$ in the SM and the MSSM).  

It is well known that in the decoupling limit, i.e.\ for $\MA \gg \MZ$, the
light $\cp$-even MSSM Higgs boson behaves in an SM-like fashion. This feature
can clearly be seen in \reffi{fig:mhm_sigmah0}, where for $\MA \gsim 150\gev$
the MSSM cross section including QCD corrections and fermion / sfermion loop
contributions is very close to the corresponding SM cross section (incorporating
QCD corrections and fermion loop contributions). As in the SM case, the
incorporated corrections are at the level of $-6\%$ in this region. For small
$\MA$, on the other hand, the couplings of the light $\cp$-even Higgs deviate
significantly from the SM case, giving rise to a suppression of the WBF
production of $h$ (while production of the heavy $\cp$-even Higgs boson, $H$,
becomes relevant in this region, see below). The relative size of the loop
corrections is much larger in this region compared to the decoupling region,
exceeding $-11\%$ for small $\MA$.

\begin{figure}[htb!]
     \subfigure[Total cross section in the MSSM, incorporating different
kinds of corrections, compared with the complete one-loop result 
in the SM.]{
          \resizebox{0.445\hsize}{!}{\includegraphics*{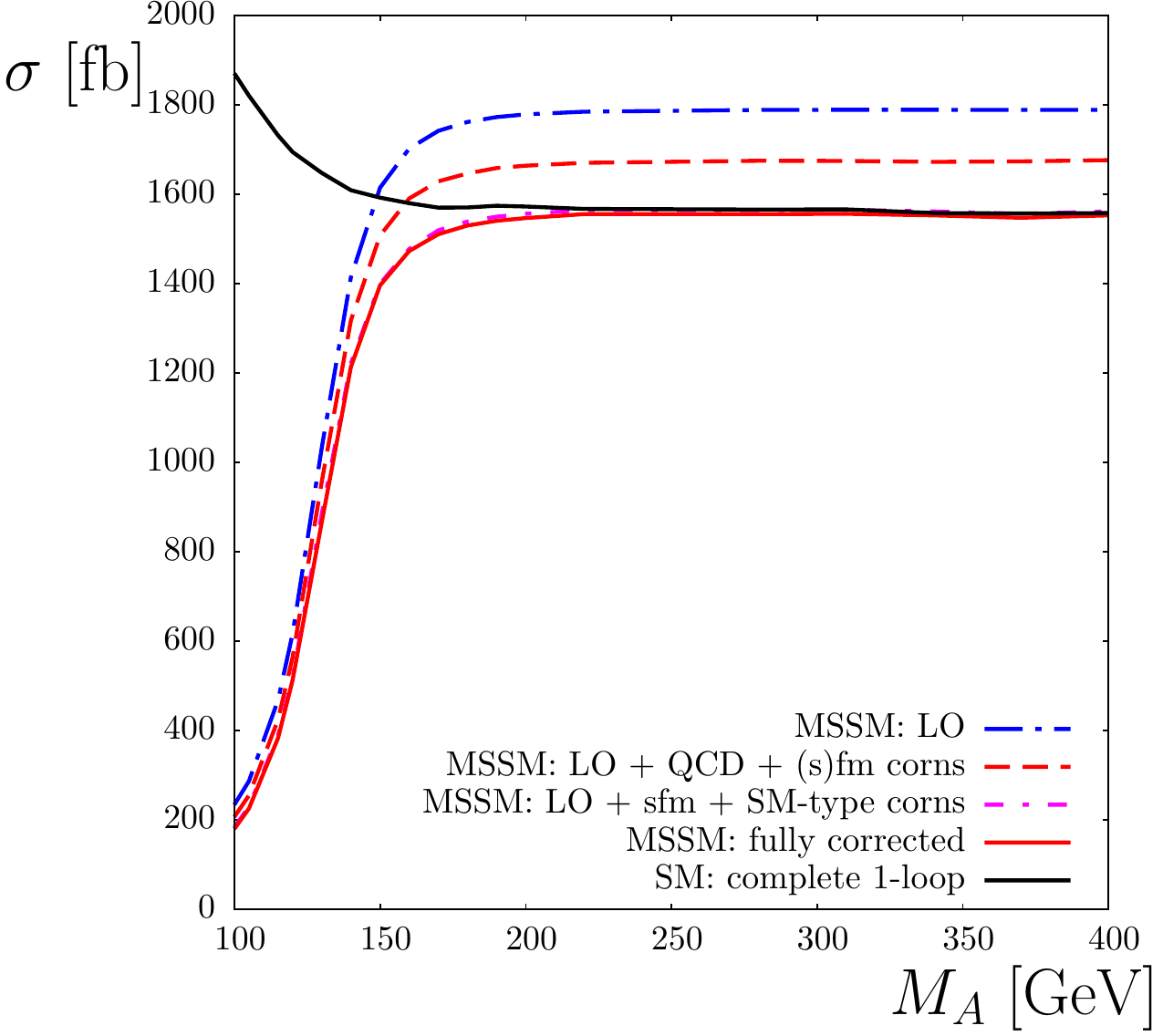}}
          }
     \hspace{.3cm}
     \subfigure[Percentage loop correction.]{
          \resizebox{0.45\hsize}{!}{\includegraphics*{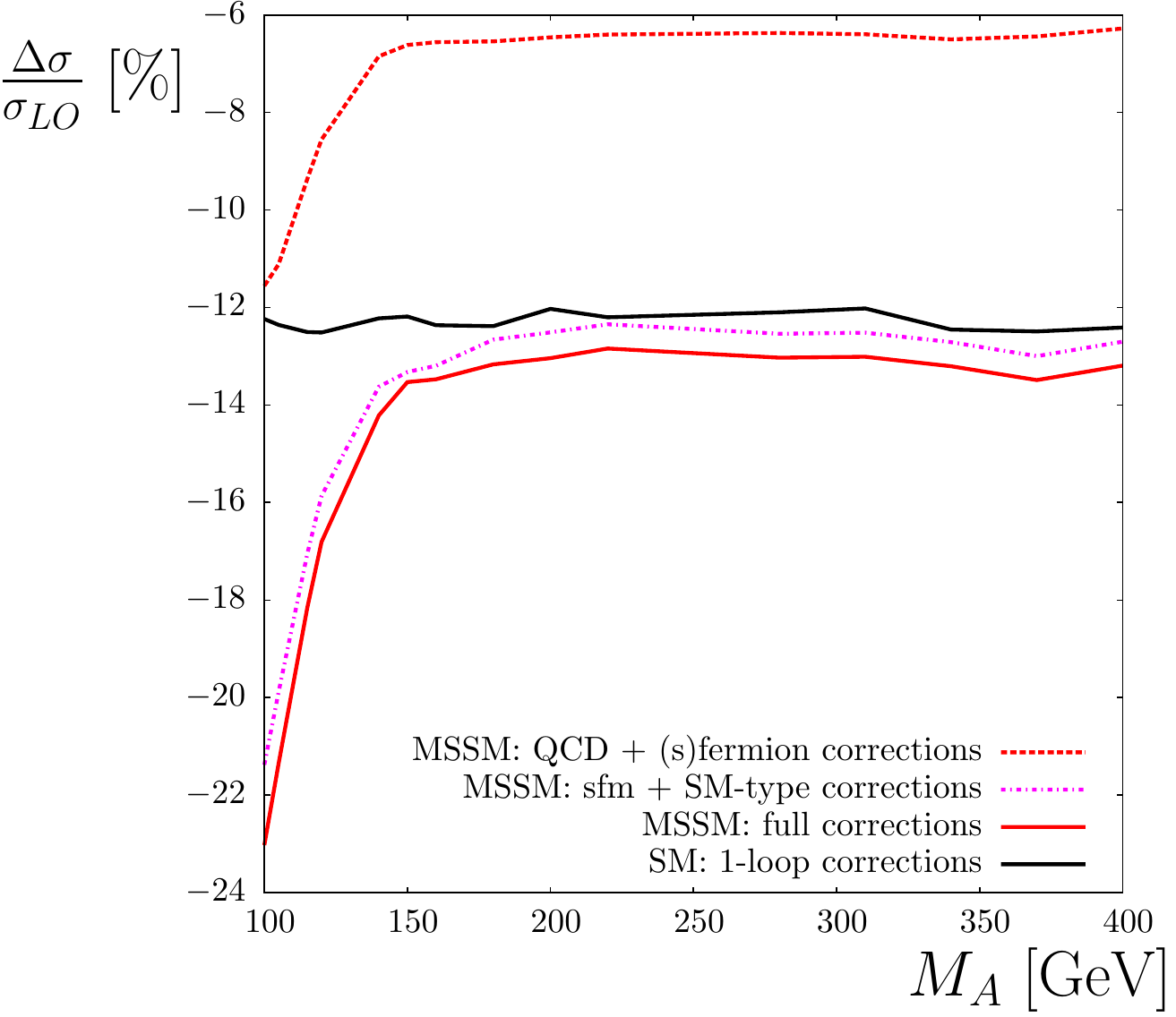}}
          } 
  \vspace{-0.4cm}
     \caption{Light Higgs boson $h$ production as a function of $\MA$ in the
\Mhmax\ scenario, with tan$\beta$ = 10. The most complete MSSM result (``MSSM:
fully corrected'') incorporates all one-loop SM-type corrections as well as all
sfermion loop contributions and the further MSSM corrections to the $VVh$, $VV$
and $qqV$ contributions. The leading order (LO) result and results containing
different parts of the higher-order corrections are also shown. The numerical
uncertainties of the Monte Carlo integration on the corrected cross section are
at the few per-mille level.}
     \label{fig:mhm_complete}
\end{figure}
Moving beyond the electroweak corrections from fermion and sfermion loops, we
now present our most complete prediction for WBF Higgs production in the MSSM.
We incorporate all SM-type corrections, i.e.\ the NLO QCD corrections already
present in {\tt VBFNLO} together with the self-energy, vertex, box and pentagon
contributions involving the gauge bosons, leptons, quarks and the particles of
the MSSM Higgs sector, as well as the real photon radiation.\footnote{As
discussed above, the SM-type contributions beyond the fermion loops play a
significant role in this context.}  Since we treat the external quarks of the
WBF process as massless, no Higgs or Goldstone bosons appear in the loops of the
box and pentagon contributions, so that those contributions are the same as in
the SM case, except for the modified coupling of the outgoing Higgs boson.
Those SM-type corrections are combined with the sfermion loop contributions in
the MSSM. The corresponding result is shown for WBF production of the light
$\cp$-even Higgs boson in \reffi{fig:mhm_complete}, labelled as
``MSSM: LO $+$ sfermion $+$ SM-type corrections''. For illustration, we also
show the prediction where furthermore the full MSSM corrections to the $VVh$,
$VV$ and $qqV$ contributions are taken into account, labelled ``MSSM: fully
corrected''. Accordingly, the ``MSSM: fully corrected'' result differs from the
complete one-loop result in the MSSM only in that we neglect the SUSY QCD (i.e.\
gluino-exchange) contributions as well as contributions from charginos and
neutralinos to the box and pentagon corrections.\footnote{As above, the leading
order cross section is parameterised in terms of the Fermi constant $G_{F}$. For
the evaluation of the quantity $\Delta r$ in the MSSM we neglect contributions
from charginos and neutralinos.}  The MSSM results are compared with the
complete one-loop result in the SM for the corresponding value of the Higgs
mass. Furthermore, the leading order result in the MSSM and the result
incorporating QCD and fermion / sfermion loop corrections, both already shown in
\reffi{fig:mhm_sigmah0}, are also displayed. 

One can see in \reffi{fig:mhm_complete} that the result including the full
SM-type corrections as well as the sfermion loop contributions (``MSSM: LO $+$
sfermion $+$ SM-type corrections'') is very close to the one where the remaining
MSSM one-loop corrections to the $VVh$, $VV$ and $qqV$ contributions are also
taken into account (``MSSM: fully corrected'').  In fact, the contribution from
charginos and neutralinos amounts to a correction of only $\sim 0.3\%$ in the
decoupling regime, and is slightly larger in the non-decoupling regime.  It
seems reasonable to expect that, as for the $VVh$, $VV$ and $qqV$ corrections,
the contribution from charginos and neutralinos will have a relatively small
effect on the boxes and pentagons.  The calculation and implementation of the
full result in the MSSM (including also the SUSY QCD contributions) will be
presented in a forthcoming publication, but the result presented here should
serve as a good approximation to the complete one-loop result in the MSSM
(supplemented by higher-order propagator-type contributions), except possibly in
parameter regions with a rather light gluino.

In the right plot of \reffi{fig:mhm_complete} the relative effect of the various
contributions is shown. The SM-type contributions beyond the QCD and fermion /
sfermion loop corrections can be seen to give rise to a downward shift of the
cross section by about $-6\%$ in the decoupling region ($\MA \gg \MZ$). As
before, the MSSM result for the light $\cp$-even Higgs boson in the decoupling
limit is found to converge to the SM result with the corresponding value of the
Higgs mass. The deviation between the relative corrections in the SM and in the
MSSM, indicating the impact of the additional SUSY loop contributions present in
the MSSM, is at the level of $0.8\%$ in this region. In the non-decoupling
regime, on the other hand, the loop effects in the MSSM can differ from those in
the SM by more than 10\% (it should be noted, of course, that the relative size
of the loop contributions is largest in the parameter region where the
production cross section for the light $\cp$-even MSSM Higgs boson is most
heavily suppressed as compared to the SM case).

\begin{figure}[htb!]
\begin{center}
          \resizebox{0.7\hsize}{!}{\includegraphics*{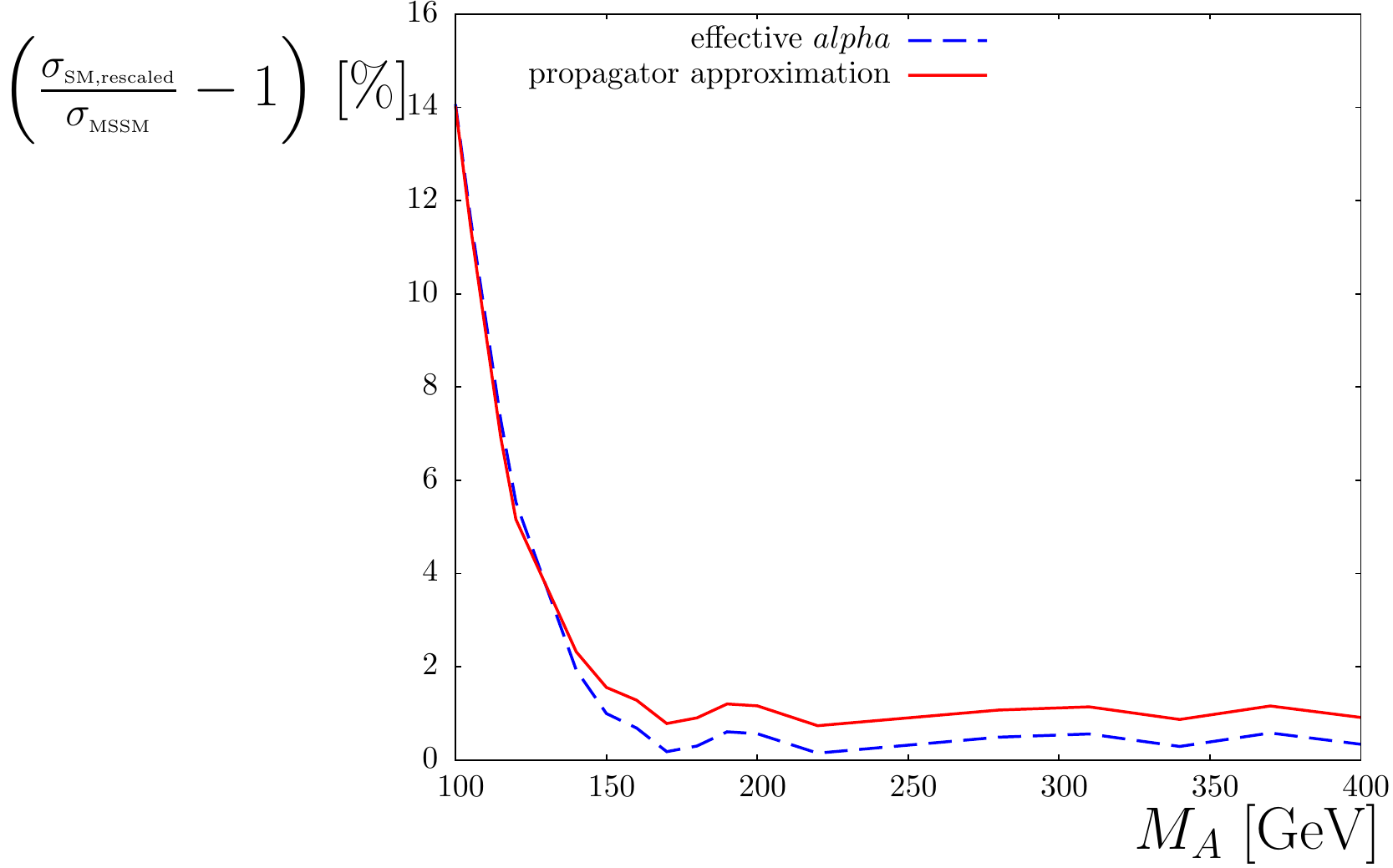}}
\end{center}
  \vspace{-1cm}
     \caption{Comparison between the most complete MSSM result for $h$
production with predictions where the complete one-loop result in the SM has
been rescaled by the propagator-type corrections in the MSSM (solid red line)
and by the effective coupling factor $\sin^{2}(\beta-\aeff)$ (dashed blue line).
The parameters are the same as in \reffi{fig:mhm_complete}.} \label{fig:rescale}
\end{figure}

An approximate treatment widely used in the literature for obtaining
MSSM predictions for Higgs production cross sections is to supplement
the loop-corrected cross section for Higgs production in the SM with an
appropriate scaling factor. For weak boson fusion, a possible scaling
factor is obtained from the Higgs propagator-type contributions
(written here for the $\cp$-conserving case; a generalisation to the
case where all three neutral MSSM Higgs bosons mix with each other is
easily possible)
\begin{equation}
 \sigma_{\rm MSSM} \sim | \sin(\beta-\alpha_{\rm tree}) Z_{hh} + \cos
(\beta-\alpha_{\rm tree}) Z_{hH} |^{2} \mbox{ } \sigma_{\rm SM} .
\end{equation}
Approximating the wave function normalisation factors further (see for instance
\citere{0003022}) leads to a simple effective coupling factor which rescales the
SM cross section,
\begin{equation}
\sigma_{\rm MSSM} \sim \sin^{2}(\beta-\aeff) \mbox{ } \sigma_{\rm SM} .
\end{equation}
In \reffi{fig:rescale} we consider the approximation where the complete one-loop
result in the SM is rescaled as described above (labelled ``propagator
approximation'' and ``effective alpha'', respectively) and compare the resulting
prediction with our most complete MSSM result as given in
\reffi{fig:mhm_complete}. As expected, in the decoupling region, $\MA \gg \MZ$,
where $h$ becomes SM-like, the simple rescaling of the loop-corrected SM result
provides a good approximation of the MSSM prediction (it turns out that in this
particular scenario the SM result scaled with $\sin^{2}(\beta-\aeff)$, which
involves additional approximations, happens to be closer to the most complete
MSSM result than for the case where the scaling factor based on the Higgs
propagator contributions is used). On the other hand, for lower $\MA$ we find
significant deviations of up to $\sim 15 \%$. 

\begin{figure}[htb!]
     \subfigure[Partonic $h$ and $Z$ production cross sections including 
(s)fermionic corrections as a function of $\tan \beta$.]{
         \label{fig:mhm_h_part}
          \resizebox{0.45\hsize}{!}{\includegraphics*{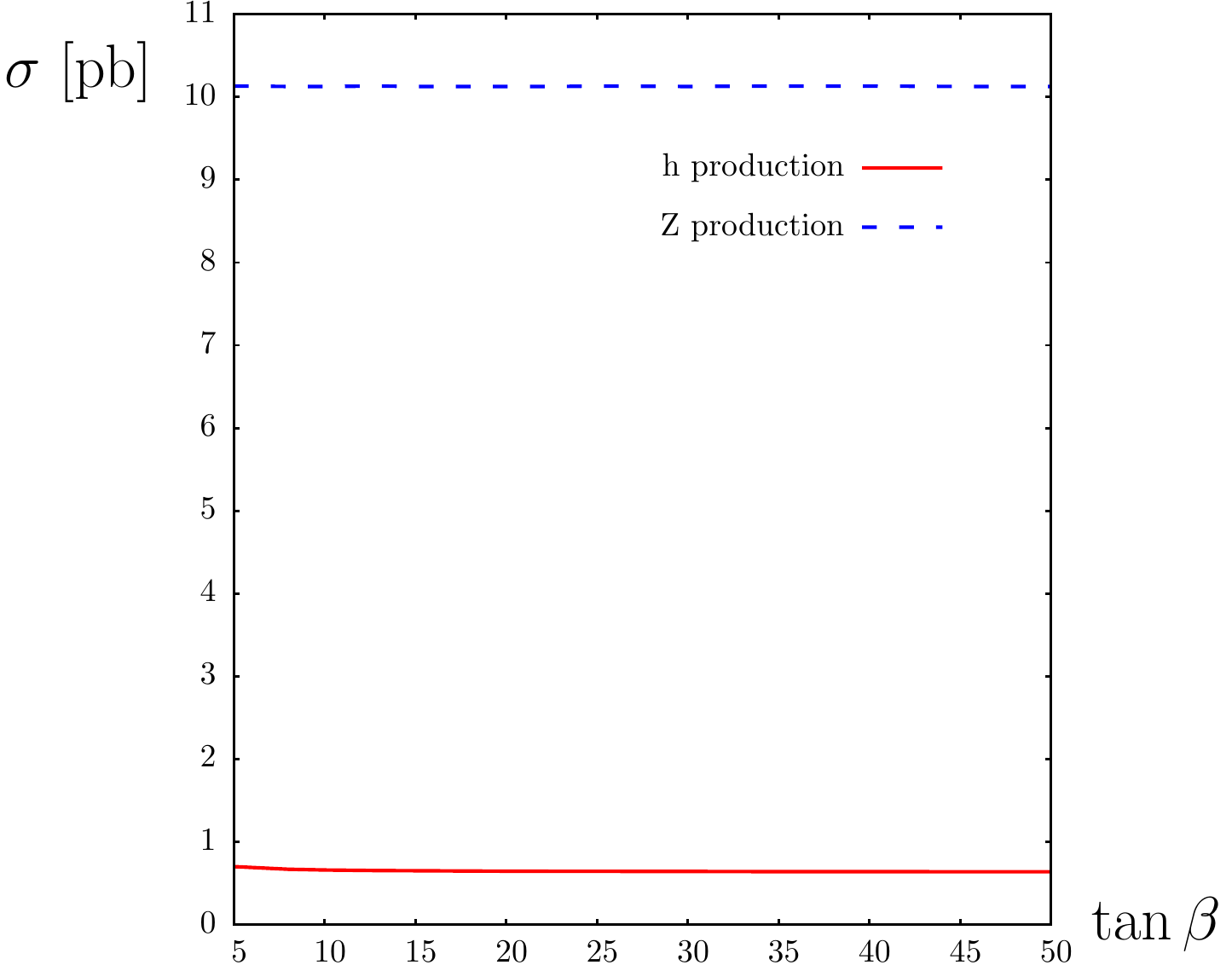}}
         }
     \hspace{.3cm}
     \subfigure[Effect of the fermionic and sfermionic loop corrections 
relative to the leading order cross sections for $h$ and $Z$ production.]{
         \label{fig:mhm_h_partF}
          \resizebox{0.45\hsize}{!}{\includegraphics*{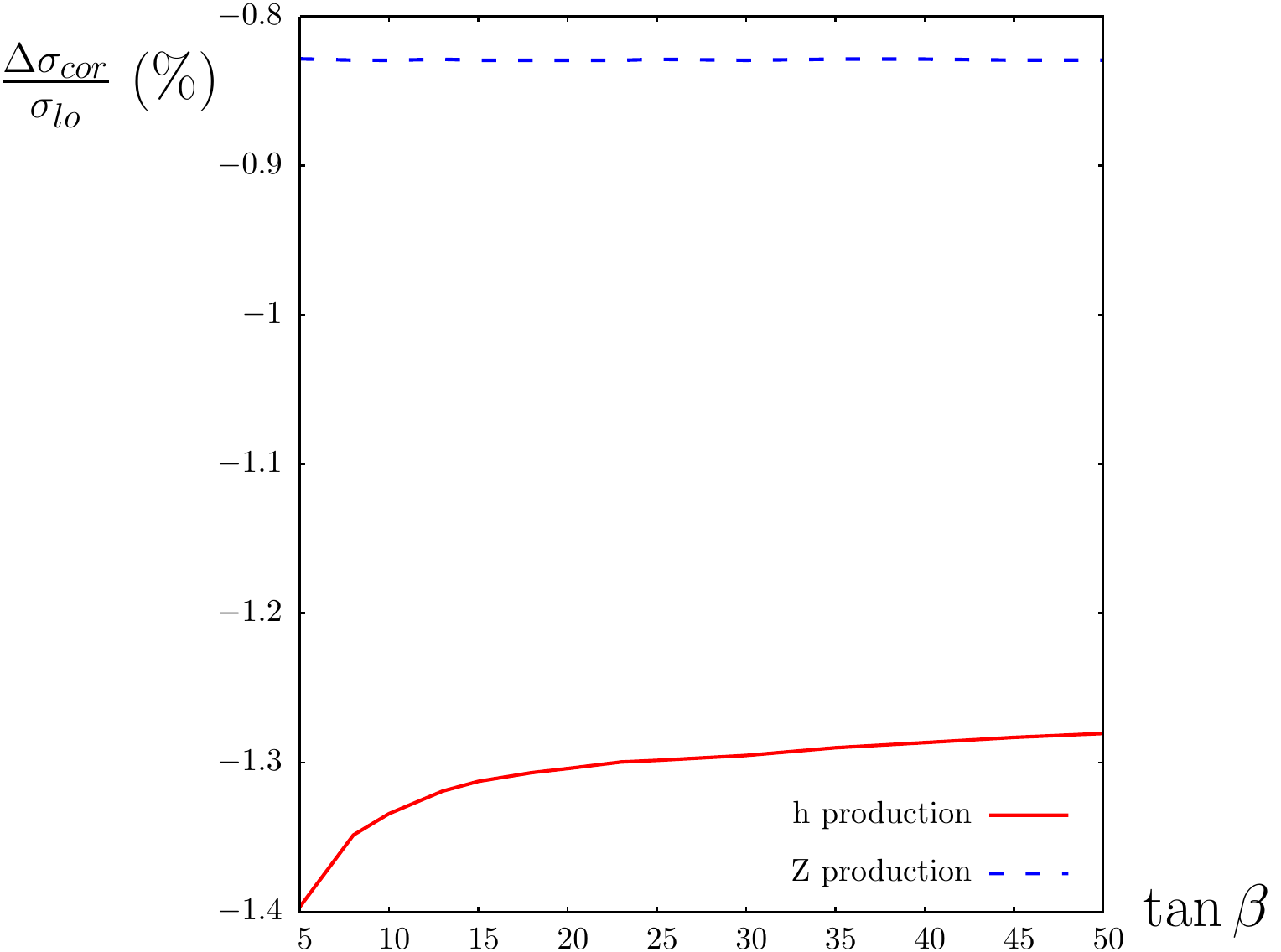}}
          }
  \vspace{-0.5cm}
     \caption{Partonic cross sections for $h$ and $Z$ production in the
\Mhmax\ scenario incorporating (s)fermionic corrections, with 
$\MA = 150\gev$ at $\sqrt{\hat{s}} = 500\gev$.}
     \label{fig:mhm_partonic}
\end{figure}

\paragraph{}
As discussed above, we have also calculated the fermion and sfermion loop
corrections to the process where a $Z$ boson is produced in WBF, which in
principle could be used as a reference process to which the Higgs production
channel could be calibrated. For simplicity, we compare our predictions for the
two processes at the partonic level, for $\sqrt{\hat{s}} = 500\gev$.
\reffi{fig:mhm_partonic} shows the results for the dominant partonic processes 
\begin{equation}
u + d \longrightarrow d + h/Z + u , \nonumber
\end{equation}
where (s)fermionic loop corrections are included. The partonic cross section for
$Z$ boson production is larger than that for $h$ production, by a factor of
$\sim 10$ (this is also the case in the Standard Model).  The loop corrections
for the two processes act in the same direction, leading to a slight reduction
of the respective cross sections. The loop corrections are at the percent level,
where the effects on the light Higgs production cross section are somewhat
larger, and (as expected) the light Higgs production cross section is more
sensitive to the parameters $M_{A}$ and $\tan \beta$. 

 \begin{figure}[htb!]
  \subfigure[A comparison of production of the light and heavy $\cp$-even Higgs
bosons as a function of $M_{A}$.]{
         \resizebox{0.46\hsize}{!}{\includegraphics*{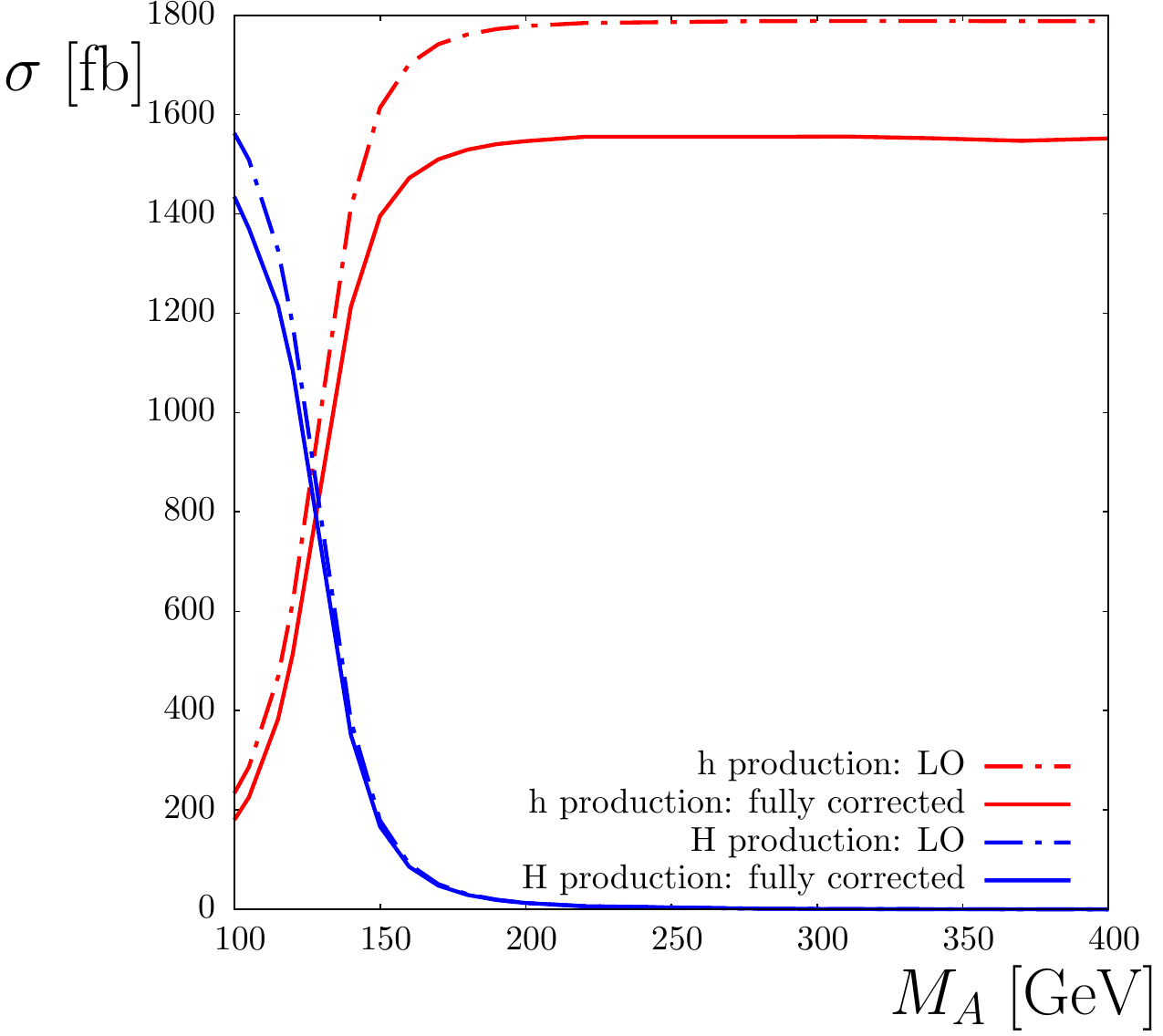}}
          } 
  \hspace {0.3cm}
  \subfigure[Loop correction percentages as a function of $M_{A}$.]{
         \resizebox{0.45\hsize}{!}{\includegraphics*{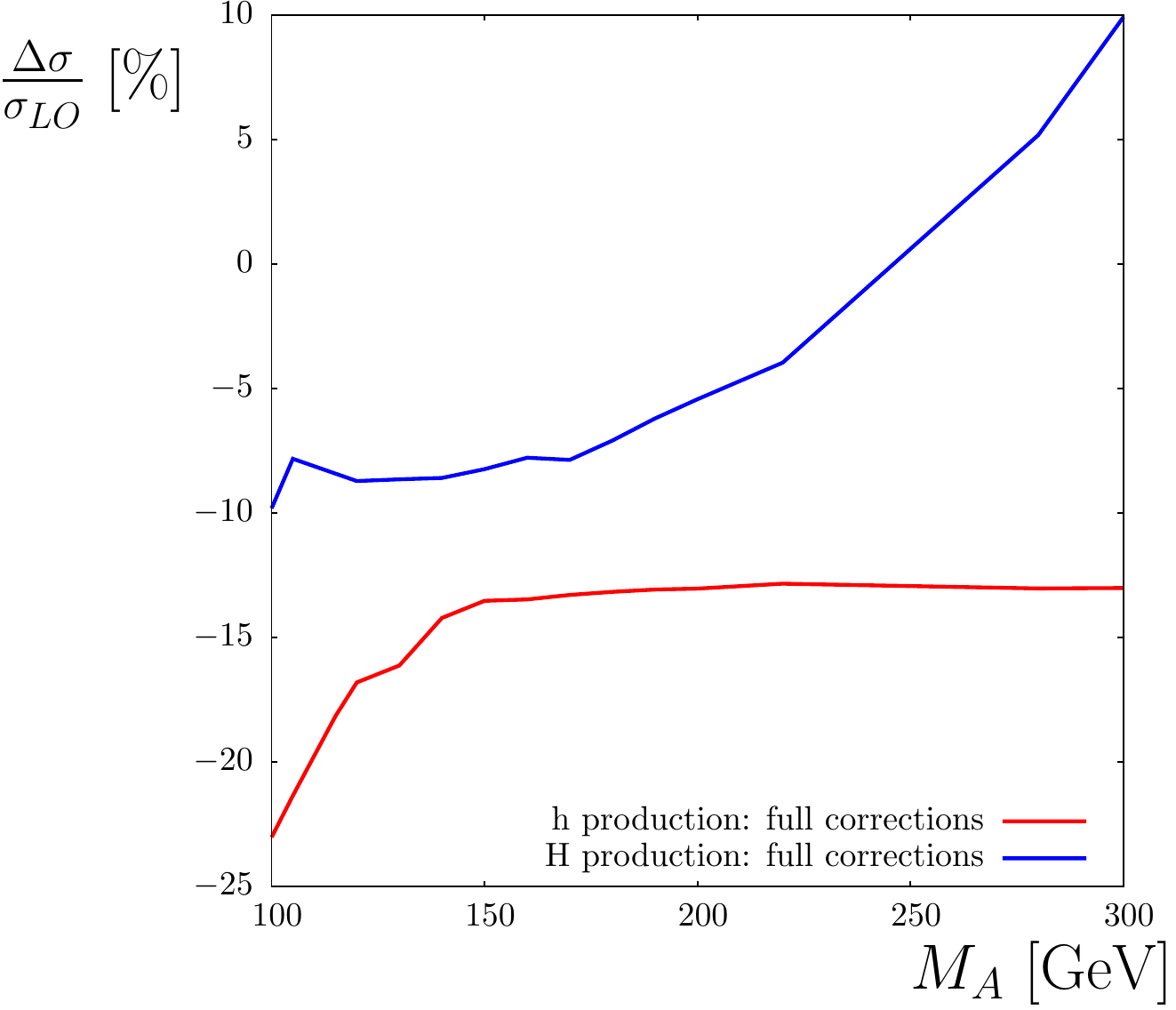}}
          } 
  \vspace{-0.5cm}
     \caption{Production of the light and heavy $\cp$-even Higgs bosons in the
\Mhmax\ scenario, with $\tan\beta$ = 10.}
     \label{fig:mhm_h0HH}
\end{figure}

\paragraph{}
While up to now we have concentrated on the production of the light MSSM Higgs
boson in WBF, we now compare the cross sections for production of the light,
$h$, and the heavy, $H$, $\cp$-even Higgs bosons of the MSSM.
\reffi{fig:mhm_h0HH} shows a comparison of the production cross sections in the
\Mhmax\ scenario of the light and the heavy $\cp$-even Higgs, as a function of
the mass of the $\cp$-odd Higgs, $\MA$, with $\tan \beta = 10$, where our
`fullest' corrections have been included (i.e.\ we neglect only the SUSY-QCD
contributions and the box and pentagon type contributions from charginos and
neutralinos).  At low values of $M_{A}$, in the non-decoupling regime where the
heavy $\cp$-even Higgs boson has SM-type couplings to gauge bosons, production
of the heavy Higgs is the dominant process.  This cross section rapidly
decreases with increasing $M_{A}$, and becomes close to zero in the decoupling
regime where the light Higgs $h$ becomes SM-like.  Due to this strong
suppression of the leading order cross section, the percentage corrections to
heavy Higgs production increase in the decoupling regime, although
the total cross section is still, of course, relatively small\footnote{Note that
the right hand plot of \reffi{fig:mhm_h0HH} only presents percentage loop
corrections for the range $M_{A} = 100$--$300\gev$, where the production cross
section of the heavy Higgs is non-negligible.}. For the heavy Higgs the loop
corrections in the \Mhmax\ scenario tend to increase the cross section in the
decoupling region (see \citere{0211204} for a discussion of scenarios where a
much larger enhancement of the heavy Higgs production cross section is
possible).  For small values of $M_{A}$ the corrections to the heavy Higgs
production cross section reach approximately $-10\%$.

 \begin{figure}[htb!]
  \subfigure[Distribution in the \Mhmax\ scenario for $h$ production, with
$M_{A} = 400 \gev$ and $M_{h} = 129.6 \gev$.]{
         \resizebox{0.45\hsize}{!}{\includegraphics*{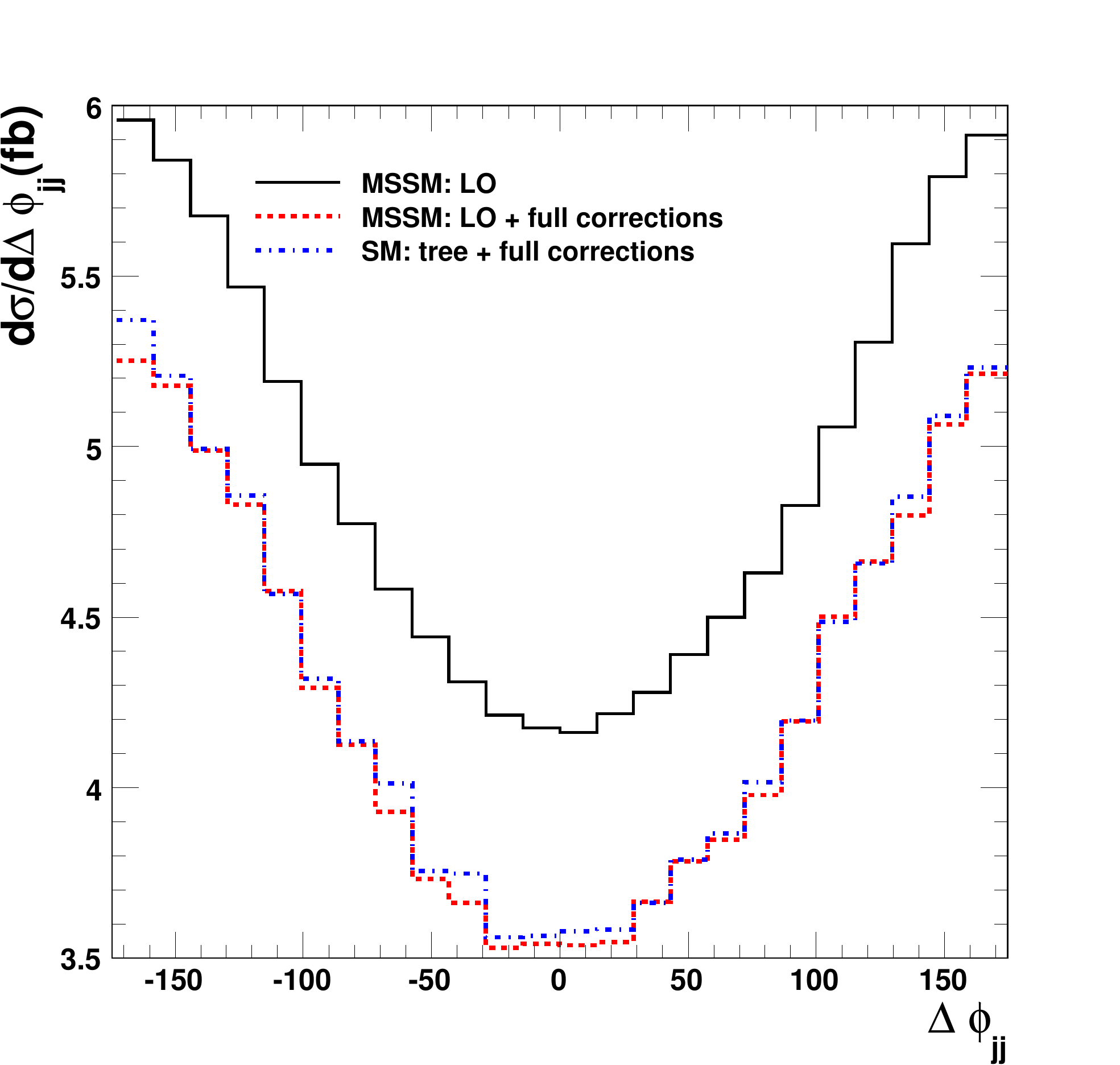}}
          } 
  \hspace{0.3cm}
  \subfigure[Distribution in the \Mhmax\ scenario with $M_{A} = 100\gev$ for the
light $\cp$-even Higgs boson ($M_{h} = 97.6 \gev$), the heavy $\cp$-even Higgs 
($M_{H} = 133.1\gev$) and Standard Model Higgs bosons ($M_{H^{SM}} = 97.6\gev$
and $M_{H^{SM}} = 133.1\gev$) with our most complete corrections included.]{
         \resizebox{0.45\hsize}{!}{\includegraphics*{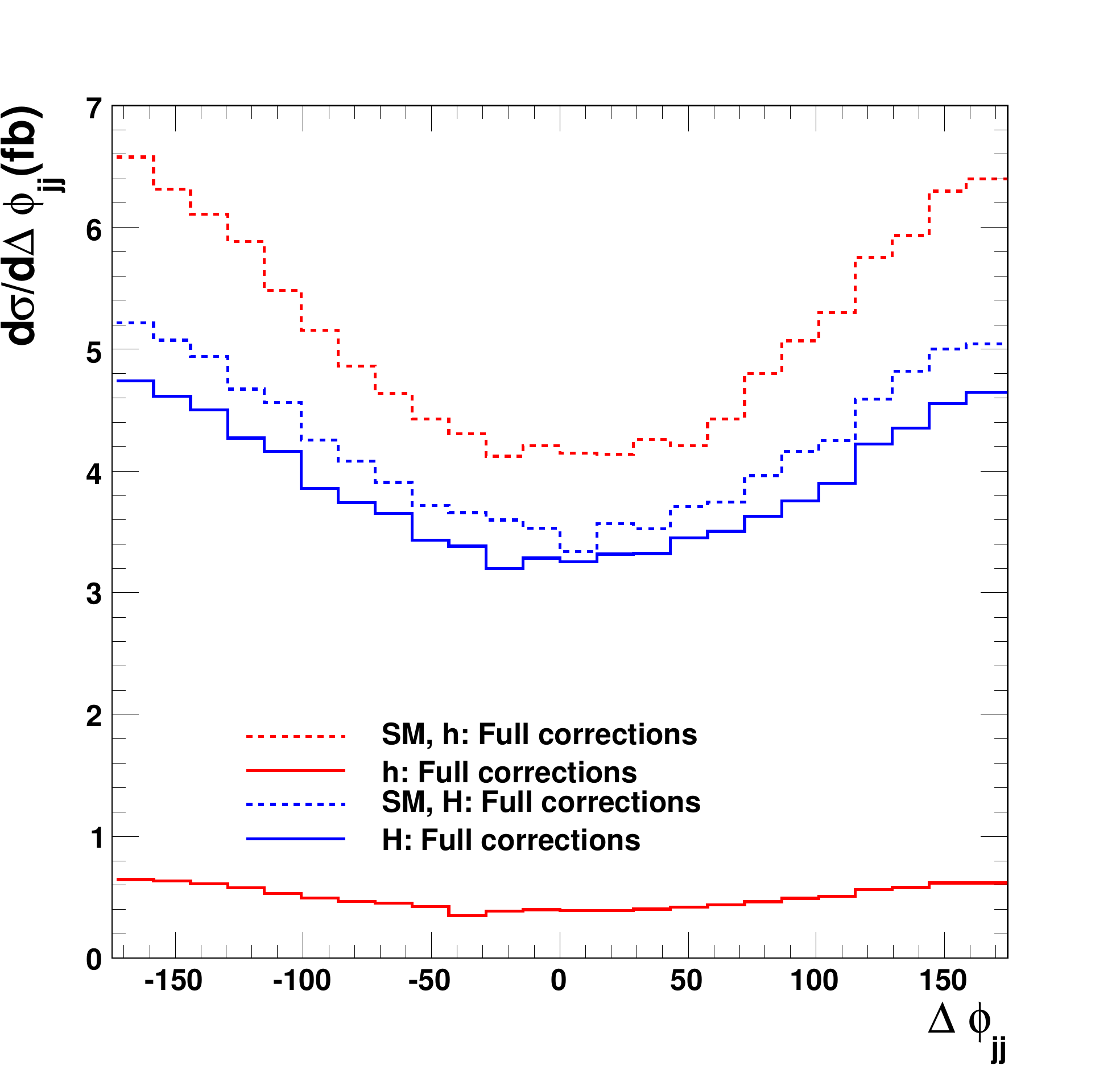}}
          } 
  \vspace{-0.5cm}
     \caption{Azimuthal angle distributions in the \Mhmax\ scenario, with
$\tan\beta$ = 10.}
     \label{fig:mhm_phi}
\end{figure}

As an example of a differential distribution, \reffi{fig:mhm_phi}(a) shows the
azimuthal angle distribution for $h$ production in the \Mhmax\ scenario in the
decoupling regime (with $\tan\beta$ = 10 and $M_{A} = 400\gev$) in comparison
with the corresponding result in the SM with the same value of the Higgs mass,
with our most complete corrections included (i.e.\ in the SM the complete
one-loop corrections are included, and in the MSSM only the SUSY-QCD and box
and pentagon contributions from charginos and neutralinos are neglected).  As
expected, $h$ production in the MSSM closely resembles the SM result in this
parameter region, so that only small differences occur between the SM and the
MSSM results. Moving out of the decoupling regime, the differences between the
MSSM and the SM become more significant. This can be seen in
\reffi{fig:mhm_phi}(b), which shows a comparison between the SM and the MSSM
results for the light and heavy $\cp$-even Higgs bosons in the \Mhmax\ scenario
for $M_{A} = 100\gev$, again with our fullest corrections included\footnote{The
two SM curves in \reffi{fig:mhm_phi}(b) are for SM Higgs bosons with masses
matching the light and heavy $\cp$-even Higgs bosons respectively.}. In this
non-decoupling region the cross section of the heavy $\cp$-even Higgs boson is
more SM-like than that of the lightest Higgs.  While differences in the total
rates are clearly visible in this example, the shape of the distribution (which
as discussed above contains information about the tensor structure of the
coupling between the Higgs and the weak boson pair) in \reffi{fig:mhm_phi}(b) is
not significantly altered in the MSSM as compared to the SM case.


\section{Conclusions}
\label{sec: conc}

We have evaluated higher-order corrections to weak boson fusion Higgs production
at the LHC in the SM and the MSSM. The weak boson fusion channel is expected to
be one of the most important channels for searching for Higgs bosons and for
determining the properties of possible Higgs candidates. Our results have been
implemented into the public Monte Carlo program {\tt VBFNLO}. Within the SM, a
complete one-loop result for weak boson fusion Higgs production has been
obtained by evaluating the full virtual electroweak corrections and photon
radiation and combining those contributions with the NLO QCD corrections already
present in {\tt VBFNLO}. Within the MSSM, the full one-loop SM-type corrections,
taking into account the extended Higgs sector of the MSSM, have been combined
with the dominant supersymmetric one-loop corrections from the scalar partners
of the SM fermions and with propagator-type corrections from the MSSM Higgs
sector up to the two-loop level. We have also presented a result where in
addition the remaining MSSM contributions to the vertex of the Higgs boson with
two gauge bosons, to the gauge boson self-energies and to the quark vertices are
incorporated, and we have verified that the numerical impact of the SUSY loop
contributions beyond the dominant sfermion loops is insignificant. Our results
have been obtained for the general case of the MSSM with arbitrary complex
parameters. The remaining supersymmetric contributions at the one-loop level,
namely contributions from neutralinos and charginos to boxes and pentagons as
well as the gluino-exchange contributions, are expected to have a small
numerical effect, except possibly in the region of a rather light gluino.
Results for those contributions will be presented elsewhere. Besides the weak
boson fusion Higgs production channel, we have also investigated loop
corrections from fermions and their scalar superpartners to $Z$-boson production
in weak boson fusion, which in principle could be of interest as a reference
process to which the Higgs production channel could be calibrated.

For those parts of our work where results already exist in the literature we
have performed detailed comparisons. For the SM case, we find complete agreement
with the results of \citere{07104749} within the numerical uncertainties. For
the case of the purely supersymmetric corrections to weak boson fusion with real
parameters we performed a comparison of the contributions to the Higgs vertex
with two gauge bosons both for the default settings of our code and for a
``tuned result'' where the higher-order corrections in the Higgs sector have
been treated in the same way as in \citere{Michael}. The numerical results in
\citere{Michael} are all given for parameters corresponding to the decoupling
limit of the MSSM, where the impact of loop corrections affecting the production
of the light $\cp$-even Higgs boson is expected to be small. We found that the
very small corrections at the level of a fraction of a percent reported in
\citere{Michael} are due to sizable cancellations between the universal
propagator-type corrections and the genuine vertex corrections. For our tuned
result we find good agreement with the results obtained in \citere{Michael},
within the expected uncertainties.

Within the SM, after applying the standard WBF cuts, we find that the
electroweak corrections give rise to a downward shift in the cross section of
order 5\% for a Higgs of mass 100--200 GeV. This is approximately the same size
as the QCD NLO corrections in this region of parameter space, leading to a full
NLO correction of order $-10$\%. Concerning the production of the light
$\cp$-even Higgs boson in the MSSM, the effects caused by loops involving
supersymmetric particles are generally small in the decoupling limit, as
expected. Comparison of our results for the MSSM and the SM (with the
corresponding value of the SM Higgs mass) shows that in this limit the SM result
is indeed recovered from the MSSM prediction to good accuracy. Away from the
decoupling region, on the other hand, the genuine vertex corrections in the MSSM
show a different behaviour compared to those in the SM, and loops involving
supersymmetric particles give rise to corrections in excess of 10\%. In fact,
approximating the MSSM prediction by the SM result scaled with an effective
coupling factor yields only satisfactory results in the decoupling region, while
for smaller values of $\MA$ deviations of up to about 15\% are possible. Particularly
large effects on the Higgs production cross section are possible in the
($\cp$-violating) CPX benchmark scenario. The loop corrections to the $Z$
production process in weak boson fusion are in general smaller than for Higgs
production and tend to go into the same direction.

In the non-decoupling region, the heavy $\cp$-even MSSM Higgs boson becomes more
SM-like, and the production of the heavy Higgs dominates over the production of
the light Higgs for very small $\MA$. In this region, we find corrections to
heavy Higgs production of about $-10\%$.  In the numerical examples that we have
analysed we find a partial cancellation between electroweak and QCD corrections
to the production of the heavy $\cp$-even MSSM Higgs in weak boson fusion in
this region.  For larger values of $\MA$, where heavy Higgs production becomes
suppressed, the relative corrections change sign and increase with increasing
$\MA$.

The implementation of our results in {\tt VBFNLO} provides a fast and efficient
tool for studying cross sections and differential distributions based on
state-of-the-art predictions in the SM and the MSSM including the effects of
experimental cuts. In this context our approach of parametrising the loop
contributions to the vertex of the Higgs boson with two gauge bosons in terms of
an effective coupling turned out to be a computationally very efficient way of
implementing this part of the calculation. The effective coupling correction was
combined with the full $2\rightarrow3$ matrix element including the remaining
loop contributions. The latest public version of {\tt VBFNLO} incorporates our results.  

\section{Acknowledgments}
\label{sec:thanks} We would like to thank A.~Denner, H.~Rzehak and W.~Hollik for
useful discussions. We are very grateful to M.~Rauch for assistance with the
SUSY comparisons, and to P. Gonzalez for formfactor comparisons. TF would also
like to thank the Galileo Galilei Institute for Theoretical Physics for the
hospitality and the INFN for partial support during part of this work. Work
supported in part by the European Community's Marie-Curie Research Training
Network under contract MRTN-CT-2006-035505 `Tools and Precision Calculations for
Physics Discoveries at Colliders' (HEPTOOLS) and MRTN-CT-2006-035657
`Understanding the Electroweak Symmetry Breaking and the Origin of Mass using
the First Data of ATLAS' (ARTEMIS).

\bibliography{references}

\end{document}